\documentclass{aastex}
\usepackage{emulateapj5, epsfig}
\usepackage{apjfonts}
\reversemarginpar
\makeatletter

\newenvironment{figurehere}
  {\def\@captype{figure}}
  {}
\makeatother

\newcommand{\percm}{${\rm cm^{-2}}$}
\newcommand{\kms}{${\rm km \ s^{-1}}$}

\newcommand{\nuflux}{${\rm erg \ cm^{-2} \ s^{-1} \ Hz^{-1}}$}
\newcommand{\lam}{$\lambda$}
\newcommand{\lamlam}{$\lambda \lambda$}
\newcommand{\lya}{Ly$\alpha$}
\newcommand{\lyb}{Ly$\beta$}
\newcommand{\zh}{Z97\nocite{zktg+97}}

\lefthead{TELFER ET AL.}
\righthead{EUV SPECTRAL PROPERTIES OF QSOS}

\begin{document}
\submitted{Accepted for publication in February 1 edition of the Astrophysical Journal}
\title{The Rest-Frame Extreme Ultraviolet Spectral Properties of QSOs}
\author{Randal C. Telfer\altaffilmark{1}, Wei Zheng\altaffilmark{1},
Gerard A. Kriss\altaffilmark{1,2}, and Arthur F. Davidsen\altaffilmark{1}}
\altaffiltext{1}{Center for Astrophysical Sciences, Johns Hopkins University, 
Baltimore, MD, 21218-2686}
\altaffiltext{2}{Space Telescope Science Institute, 3700 San Martin Drive, 
Baltimore, MD, 21218}

\begin{abstract}
We use a sample of 332 Hubble Space Telescope spectra of 184 QSOs with $z > 0.33$
to study the typical ultraviolet spectral properties of QSOs, with emphasis on the ionizing
continuum.  Our sample is nearly twice as large as that of \citet{zktg+97} and provides much better 
spectral coverage in the extreme ultraviolet (EUV).  The overall 
composite continuum can be described by a power law with index $\alpha_{EUV} = -1.76 \pm 0.12$
($f_\nu \propto \nu ^\alpha$) between 500 and 1200 \AA.  The corresponding results for
subsamples of radio-quiet and radio-loud QSOs are $\alpha_{EUV} = -1.57 \pm 0.17$ and
$\alpha_{EUV} = -1.96 \pm 0.12$, respectively.  
We also derive $\alpha_{EUV}$ for as many individual objects in our sample as possible,
totaling 39 radio-quiet and 40 radio-loud QSOs.  The typical individually measured values
of $\alpha_{EUV}$ are in good agreement with the composites.
We find no evidence for evolution of $\alpha_{EUV}$ with redshift for either radio-loud or
radio-quiet QSOs.  However, we do find marginal evidence for a trend towards harder
EUV spectra with increasing luminosity for radio-loud objects.
An extrapolation of our radio-quiet QSO spectrum is consistent with existing
X-ray data, suggesting that the ionizing continuum may be represented by a single power law.
The resulting spectrum is roughly in agreement with models of the intergalactic medium
photoionized by the integrated radiation from QSOs.
\end{abstract}


\section{INTRODUCTION\label{sec:intro}}
One of the most striking characteristics of QSO spectra is their similarity, even
over a span of a few decades in luminosity and a large range in redshift.  
This similarity in
spectral properties makes the production of composite spectra an appealing and useful tool 
for studying the general properties of the QSO population.  In recent years, 
the number of known quasars has increased dramatically due to comprehensive
surveys and, in particular, the effective use of fiber-optic spectrographs.
Very high signal-to-noise ratio (S/N) composites consisting of hundreds
to thousands of optical spectra have been created out of spectral data acquired for various
surveys, including composites from the Large Bright Quasar Survey \citep[LBQS;][]{fhfc+91},
the FIRST survey \citep{btbg+00}, and the Sloan Digital Sky Survey \citep[SDSS;][]{vand+01}.
Composites of subgroups of QSOs have also been generated to study the dependence of the
spectra on particular properties, such as radio brightness 
(\citealt*{fhi93}; \citealt{btbg+00}),
X-ray brightness \citep{gree98}, and radio morphology \citep{bahu95}.  The ultimately
enormous number of QSO spectra from the SDSS promises the ability to study in detail the
spectral dependence on various properties, including luminosity and
redshift.

These studies have contributed greatly to our detailed knowledge of QSO spectral properties
at wavelengths longer than \lya.  However, studies performed with optical spectra are
limited in the amount of information they can supply at wavelengths shorter than \lya.
These objects can only be observed below \lya\ for $z \gtrsim 2$ from the ground, and in
practice the limit from survey spectra is even more restrictive.  For example, the SDSS
spectroscopic data begin at $\sim 4000$ \AA, and thus \lya\ can only be observed for
$z > 2.3$.  At such high redshift, the \lya\ forest absorption is extremely dense, 
depressing the flux and greatly reducing the S/N.  Exploring the spectral properties
of QSOs below \lya\ and deeper into the extreme ultraviolet (EUV) thus requires the use of 
ultraviolet spectra.

The EUV range is important for several reasons: (1) The photons below
912 \AA\ are believed to be the main sources of line formation via 
photoionization and they should be related to the emission lines seen in the
near ultraviolet (NUV) and optical regions; (2) The continuum shape in this range helps define the
peak in the QSO energy output in the UV known as the Big Blue Bump, which may be 
relevant to the soft X-ray fluxes. Evidence suggests that the EUV continuum is related to the 
soft X-ray, which may help to explain the origin of the Big Blue Bump; (3) 
EUV emission from QSOs is thought to be an important, if not dominant, source of ionization
for the intergalactic medium (IGM), and the physical state of the IGM can depend on the 
exact shape of the ionizing continuum.

The optical composites suggest that the continuum can be represented by a power law 
between $\sim 1200$ and 5000 \AA, with a power-law index $\alpha$ around $-0.3$ to $-0.5$
($F_{\nu} \propto \nu^\alpha$).  However, \citet[hereafter Z97]{zktg+97} find that there
is a distinct change in the spectral shape in the EUV.
Using the Hubble Space Telescope (HST) Faint Object Spectrograph (FOS) spectra of 101 quasars 
with $z>0.33$, \zh\ produced a composite spectrum that 
covers a range between 350 and 3000 \AA.  The resulting composite indicated a break in the 
continuum shape around 1050 \AA. The EUV continuum shape for the \zh\ radio-quiet composite
could be approximated with a power law of index $-1.8$. By comparison, \citet{lfew+97}
derived a mean soft X-ray power-law index of $-1.72$ for a sample of $z < 0.4$ radio-quiet QSOs.  
The similarity of the power-law indices in these two independent studies suggests 
a common origin for the UV / X-ray continuum.  Thus, although the full ionizing continuum of
QSOs is not directly observable due to the opacity of the Galaxy, by studying a significant
portion of the EUV continuum we may plausibly gain knowledge of the ionizing spectrum as
a whole.

In this paper, we extend the work of \zh\ to include the full sample of $z>0.33$ QSO
observations with the HST FOS and Goddard High Resolution Spectrograph (GHRS), as well as 
available Space Telescope Imaging Spectrograph (STIS) data.
We begin in \S\ref{sec:obs} by describing the sample, our selection criteria, and the
details of our processing of the data.  We proceed in \S\ref{sec:combine} by describing
the technique for creating our composites.  In \S\ref{sec:results} we first discuss our results
and the corresponding uncertainties.  We then discuss how we estimate luminosities using
the composite for the purpose of characterizing the sample.  This is followed by a
presentation of our radio-loud and radio-quiet composites, and finally a discussion of
the broad emission-line properties.  In \S\ref{sec:indiv} we discuss how we characterize 
the continuum of individual objects in our sample.
We compare the results with the composite and also search for possible
correlations of the EUV continuum with luminosity and redshift.  We compare our results
with other composites in \S\ref{sec:discuss}, and then discuss the implications of our
results.  We summarize our results in \S\ref{sec:summary}.

\section{DATA\label{sec:obs}}

\subsection{The Sample\label{sec:sample}}
Our data sample consists of all known spectra of $z>0.33$ QSOs taken with the HST
FOS and GHRS.  We add to this
all QSO spectra from STIS that were publicly
available as of 2000 August.  The redshift limit was chosen such that the region below the 
intrinsic Lyman limit (912 \AA) is observable by HST.  We removed spectra from the sample
for the following reasons:
\begin{enumerate}
\item{The object is a known or suspected Broad Absorption Line QSO.}
\item{The spectrum is high resolution with small wavelength coverage (medium and high 
resolution modes on GHRS and STIS).}
\item{The S/N is very low ($\lesssim 1$ per pixel).}
\item{After trimming (see \S\ref{sec:reduce}), very little spectrum remains
($\Delta\lambda / \lambda \lesssim 0.2$).}
\end{enumerate}
In addition, if two spectra are almost entirely redundant -- for example, if both
FOS G130H and GHRS G140L data are available -- then only the higher quality data are
used.  However, in cases where there is only a partial redundancy (for example, FOS
G270H and FOS G160L), both spectra are included.

After applying these selection criteria, our final sample consists of 332 spectra of
184 QSOs, nearly twice as large as the sample from \zh\footnote{\zh\ stated that their
sample consisted of 284 spectra of 101 QSOs.  However, many of these spectra were separate
observations of the same QSO using the same instrument and grating that we have since 
combined.  Counting spectra in a manner consistent with the present sample, the \zh\ sample 
consisted of 184 spectra of 101 QSOs.}.  
This final sample contains 13 spectra with FOS grating G130H, 93 with
FOS G190H, 139 with FOS G270H, 25 with FOS G400H, 2 with FOS G570H, 51 with FOS G160L, 
6 with GHRS G140L, 2 with STIS G140L, and 1 with STIS G230L.  
The redshift distribution of the corresponding
QSOs is shown in Figure~\ref{fig:reddist}, along with the corresponding distribution of \zh.

\begin{figurehere}
\centerline{\psfig{file=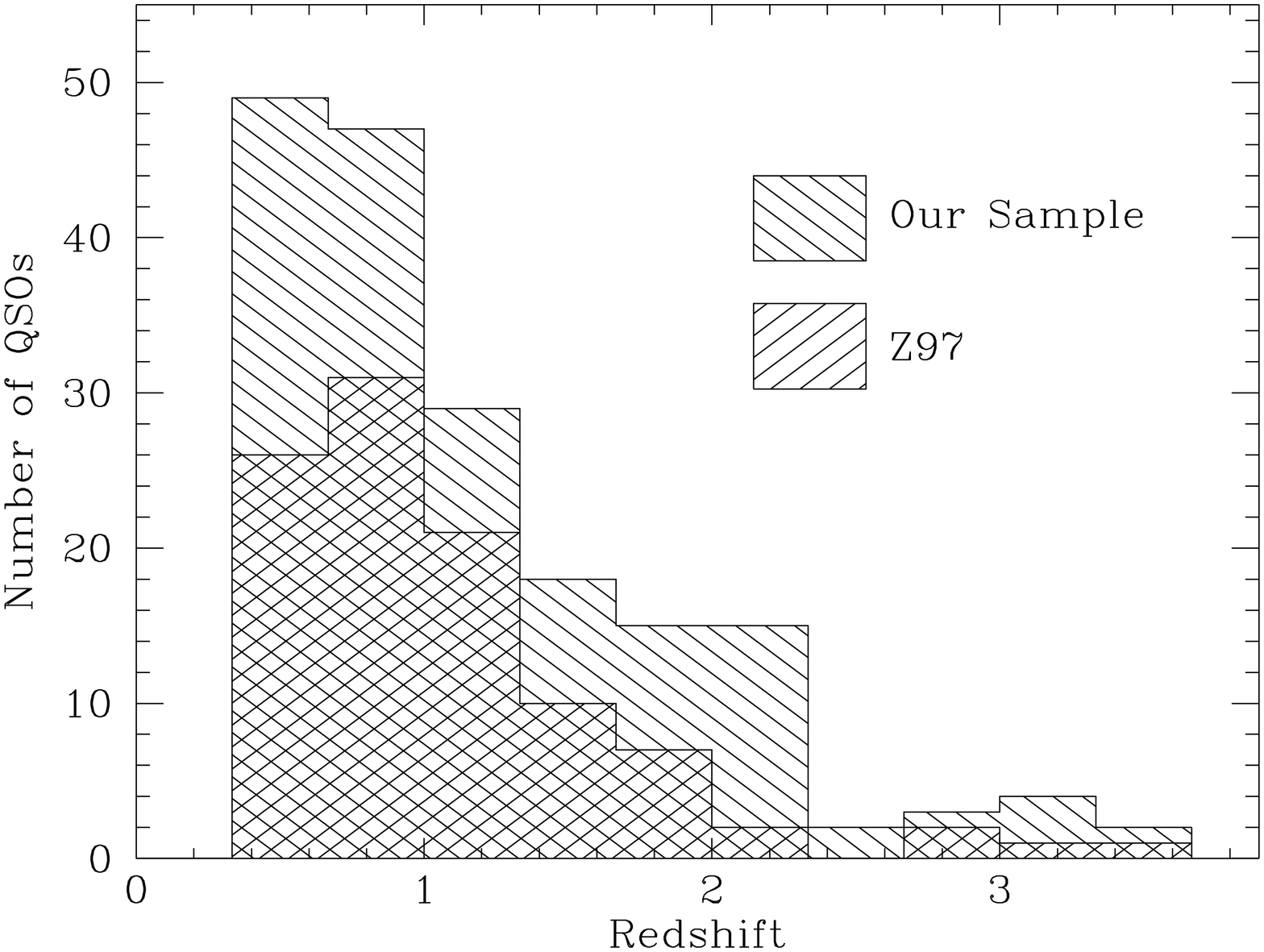,angle = -90, width=9cm}}
\caption{Redshift distribution of sample QSOs with a bin size of 0.33, comparing
the current sample with Z97.\label{fig:reddist}}
\end{figurehere}
\vspace{0.2cm}

\subsection{Reduction\label{sec:reduce}}
Prior to combination, each spectrum is processed with a series of tasks which we describe in
more detail below.  Specifically:
\begin{enumerate}
\item{The spectrum is corrected for Galactic extinction.}
\item{Strong features not intrinsic to the QSO are masked out, including instrumental 
artifacts, airglow, and strong absorption.}
\item{If there are one or more intervening Lyman limit absorbers in the sight line for
which the optical depth can be well determined and the data
below the Lyman limits meet our S/N criterion (S/N per pixel $\gtrsim 1$), we correct
for the Lyman continuum absorption of the systems.}
\item{If there is an optically thick Lyman limit absorber in the spectrum such that the
data below the Lyman edge are not recoverable, we remove the data below
the edge from the spectrum.}
\item{A statistical correction is applied for the accumulated Lyman line and Lyman limit
opacity of absorbers below our detection threshold.}
\item{The spectrum is shifted into the object rest frame.}
\item{The spectrum is resampled into common wavelength bins to facilitate the combination
process.}
\end{enumerate}
The order in which these steps are applied is unimportant, with the exceptions that the
Galactic extinction correction must be performed before the spectra are shifted into the rest
frame and the resampling must be performed after the spectra are shifted. 
We now discuss the individual steps in appropriate detail.

The Galactic extinction correction is performed according to the analytic form of 
\citet*{ccm88} using $E(\bv)$ values from \citet*{sfd98}.  We assume $R_V = 3.1$ for all 
objects.

We mask undesired data by simply flagging the individual pixels involved
so that they are ignored by the code that combines the spectra.  Some spectra contain
odd spikes that are clearly not intrinsic to the QSO, probably caused by
malfunctioning diodes not recognized by the calibration software.  These are masked out,
as are airglow lines from \lya\ and \ion{O}{1} \lam 1304 in FOS G130H, FOS G160L, and
GHRS G140L spectra.  Also, all FOS G160L data above 2300 \AA\ are removed as this region
is generally contaminated by second-order \lya\ airglow.  We search each spectrum by
eye for strong absorption features that change the apparent character of the 
spectrum and could have some influence on the final composite, including damped lines
and strong clusters of lines.

For two well-studied objects, LB~9612 and HS~1700+641, we correct for the Lyman limit
absorbers using redshifts and column densities from the literature, \citet{dfiw98} and
\citet{vore95}, respectively.  For all other objects, the individual Lyman limit systems 
are identified by eye. For most systems the redshift
is well-determined by the identification of many Lyman lines, although in the low-resolution
FOS 160L data this is generally not possible and the redshift of the system is estimated
from the position of the Lyman break itself.  The optical depth of each Lyman limit is
estimated as $\ln (F_+ / F_-)$, where $F_-$ is the median flux of the pixels just below the break,
from 890--911 \AA\ in the absorber frame, and $F_+$ is the median flux of the pixels in the
windows 933--936, 940--948, and 952--960 \AA\ in the absorber frame.  These windows are chosen to
avoid absorption by the Lyman lines.  After the correction
for the Lyman limit is performed, typically a large absorption feature remains due to
the clustering of high-order Lyman lines that is masked out as described above.
If we cannot achieve an acceptable correction because the S/N below the break is too low, 
which generally occurs for $\tau \gtrsim 2$, the portion below the break is simply removed.

We are confident that we are correcting for nearly all Lyman limit absorbers with $\tau > 0.3$,
corresponding to a neutral hydrogen column density $\log N({\rm HI}) ({\rm cm^{-2}})> 16.7$.
However, the accumulated absorption of weaker absorbers, being much more numerous,
contributes strongly to the opacity of high-redshift objects, producing a broad trough centered 
near 700 \AA\ known as the Lyman valley \citep{moja90}.  We therefore apply a 
statistical correction for these unidentified absorbers.

We characterize the distribution of Lyman forest absorbers by the empirical formula:
\begin{equation}\label{eq:valley}
\frac{\partial ^2 n}{\partial z \partial N} \propto (1+z)^\gamma N^{-\beta},
\end{equation}
where $n$ is the number of lines, $z$ is redshift, and $N$ is the column density of neutral
hydrogen.
As first pointed out by \citet{tytl87}, $\beta = 1.5$ provides a reasonable fit to the column 
density distribution of Lyman forest absorbers over the entire range of observable 
column densities ($\sim 3 \times 10^{12} - 10^{22}$ \percm).  However, known deviations from
a pure power law are large enough to have observable consequences on the Lyman valley
correction.  We adopt values from \citet{pwrc+93}:  $\beta = 1.83$ for
$3 \times 10^{14} < N < 10^{16}$ \percm\ and $\beta = 1.32$ for $N > 10^{16}$ \percm.  For 
$N < 3 \times 10^{14}$ \percm, we use $\beta = 1.46$ from \citet{hkcs+95}, down to their 
limiting column density of $3 \times 10^{12}$ \percm.  For the redshift evolution power-law 
index, we use $\gamma = 2.46$ \citep*{prs93}.  The entire distribution is normalized by
demanding that the line number be consistent with \citet{hkcs+95} at the low column
density end.  Specifically, we set 
$\partial ^2 n / \partial z \partial N = 6.76 \times 10^{-13}$ at $z=2.8$ and $N=10^{14}$ \percm.
At breaks in the index $\beta$, we demand that the distribution be continuous.  For 
calculating the line opacity, we assume a $b$ parameter of 30 \kms.

When the wavelengths are divided by $(1+z)$ to shift the spectra into the object rest frame,
the fluxes in $F_\lambda$ are multiplied by $(1+z)$. Each resulting spectrum then has the
property that when the flux is multiplied by $4 \pi D_L^2$, where $D_L$ is the luminosity
distance to the QSO, the result is the true luminosity $L_\lambda$ of the QSO.

The spectra are resampled into aligned wavelength bins of common size.  The flux in each
of the bins of a resampled spectrum is determined as the mean of the flux in the bins of
the old spectrum that overlap the new bin weighted by the extent of the overlap between
the new bin and old bin in wavelength space.  The errors per rebinned pixel are calculated 
in a manner consistent with this weighting.  More precisely, we use the following formulas
for calculating the rebinned fluxes and errors:
\begin{eqnarray}
F_r & = & \frac{\sum_i F_i \delta\lambda_i ^{\prime}}{\sum_i \delta\lambda_i ^{\prime}} \\
E_r & = & \frac{\{\sum_i [(E_i \delta\lambda_i ^{\prime})^2 (\delta\lambda_i / \delta\lambda_i ^{\prime})]\}^{1/2}}{\sum_i \delta\lambda_i ^{\prime}},
\end{eqnarray}
where $F_r$ and $E_r$ are respectively the flux and error of the rebinned pixel,
$F_i$ and $E_i$ are the fluxes and errors in the pixels of the original spectrum, 
$\delta\lambda_i$ are the sizes of the original bins, and $\delta\lambda_i ^{\prime}$
are the overlap of the old pixels with the new bin.  The term 
$\delta\lambda_i / \delta\lambda_i ^{\prime}$ in the error corrects for the fact that
the error in the flux for a fraction of a pixel scales as the $- 1/2$ power of the overlap.
This method introduces correlated errors between adjacent pixels, but we do not track the
covariant errors.

\section{COMBINATION TECHNIQUE\label{sec:combine}}
The usual method of producing composite spectra is the bootstrap.
In the bootstrap method, the spectra are included in sorted order, either from
shorter wavelength coverage to longer or vice versa.  The composite is then constructed
as follows:
\begin{enumerate}
\item{The first spectrum is defined to be the partially formed composite, perhaps with
some arbitrary renormalization for convenience.}
\item{The next spectrum is renormalized such that the mean flux of the spectrum and the 
partially formed composite are the same in the region over which the two overlap, 
excluding regions of strong emission lines.}
\item{A new composite is computed, this being the mean flux in each wavelength bin of the
contributing spectra.}
\item{Steps 2 and 3 are repeated until all spectra are included.}
\end{enumerate}

The primary source of flexibility in this method is the manner in which the spectra are
normalized.  For the normalization routine, we mask out wavelength regions with strong emission
lines; specifically, 750--800, 1000--1050, 1160--1265, 1380--1420, 1470--1610, 1830--1950, 
and 2700--2880 \AA.  The normalization of the spectra will therefore be driven by the 
continuum, which is our primary interest.  In calculating the normalization factor for
each spectrum, each pixel is weighted by the number of spectra contributing to the 
partially formed composite.  This prevents the normalization from being dominated by
regions of the composite that are determined by few objects and are therefore uncertain.
Thus, each spectrum is renormalized by multiplying the flux at each pixel by a factor
\begin{equation}
f_{\rm renorm} = \frac{\sum_i F_{c,i} N_{c,i}}{\sum_i F_{s,i} N_{c,i}},
\end{equation}
where $F_c$ is the flux in the partially formed composite, $N_c$ is the number of spectra
contributing to the composite, and $F_s$ is the flux in the spectrum to be renormalized.
The sum is over the wavelength region where the two overlap, excluding the masked 
wavelength regions.

The bootstrap method has a potential drawback in that it can be sensitive to the order in
which spectra are added to the composite (longer wavelengths to shorter or vice versa),
particularly at the beginning of the
process where only a few spectra, which may not be representative of the entire sample, are
determining the shape of the composite.  To avoid this, we use a variation on the bootstrap
method.  As is evident in Figure~\ref{fig:count}, which shows the number of spectra in our
sample as a function of wavelength, the number of spectra 
reaches a maximum in the region around \lya.  We therefore
use the spectra in this middle region to define well the central portion of the composite,
then bootstrap to the extremes.  We choose the region from 1050--1150 \AA\ as the central
window, since it is relatively free of emission lines.  We divide the overall sample into
three subsamples:  (1) the spectra that include the entire central spectral window, (2) those
at longer wavelengths, and (3) those at shorter wavelengths.  In our method, we first create
a composite of just the spectra in group (1), where we renormalize each spectrum so that
the mean flux in the central window is unity.  Those in group (2) are then bootstrapped in
sorted order to longer wavelengths as described before, then group (3) is bootstrapped to
the shortest wavelengths.  Since the central region is well defined by group (1), the
order in which group (2) and group (3) are included is unimportant.

\begin{figurehere}
\centerline{\psfig{file=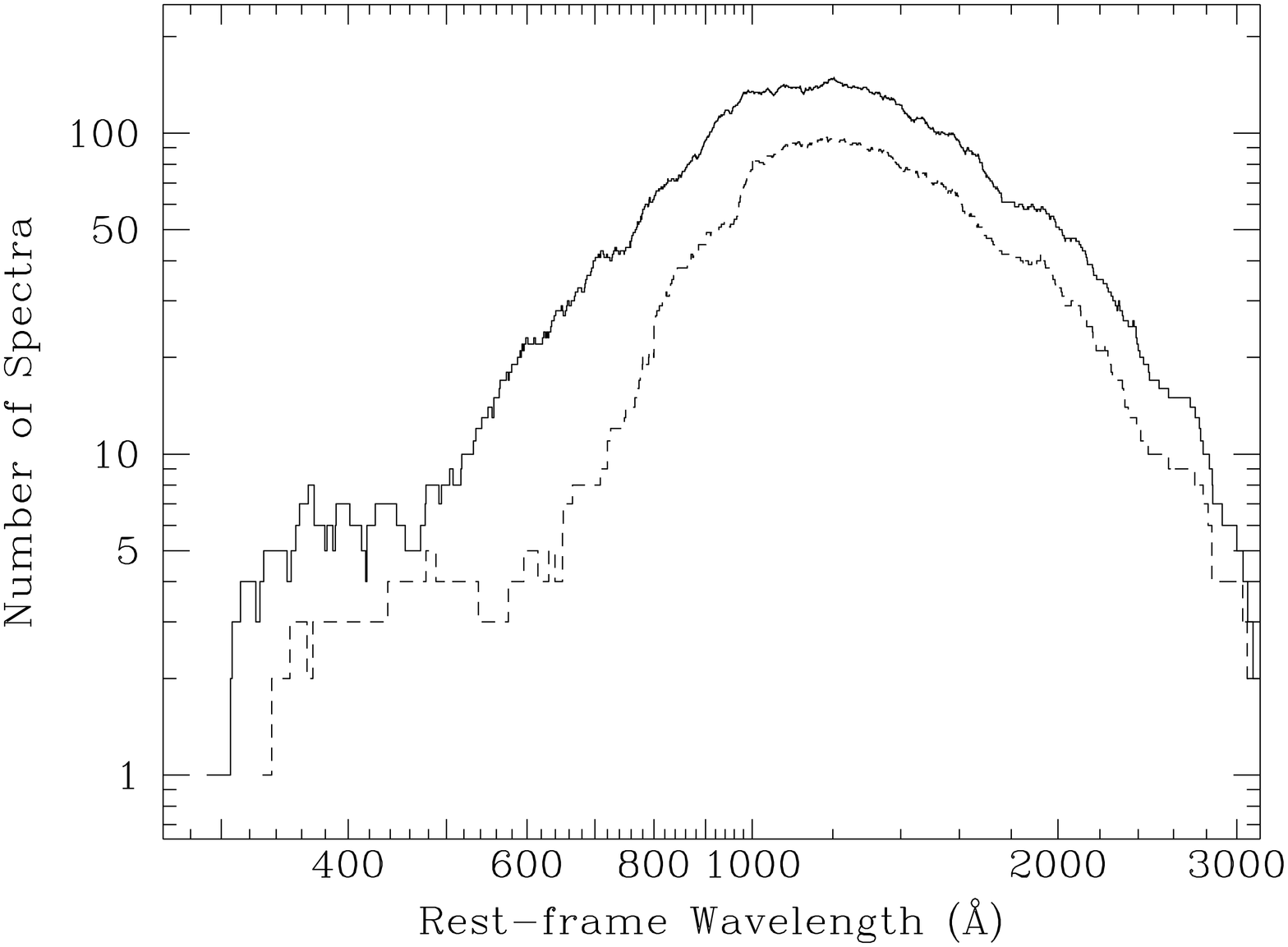,angle = -90, width=9cm}}
\caption{Number of merging spectra in the overall composite as a function
of wavelength (solid line).  The corresponding plot for Z97 (dashed line) is shown for 
comparison.\label{fig:count}}
\end{figurehere}
\vspace{0.2cm}

\section{RESULTS\label{sec:results}}

\subsection{Overall Composite\label{sec:comp}}

\begin{figure*}
\centerline{\psfig{file=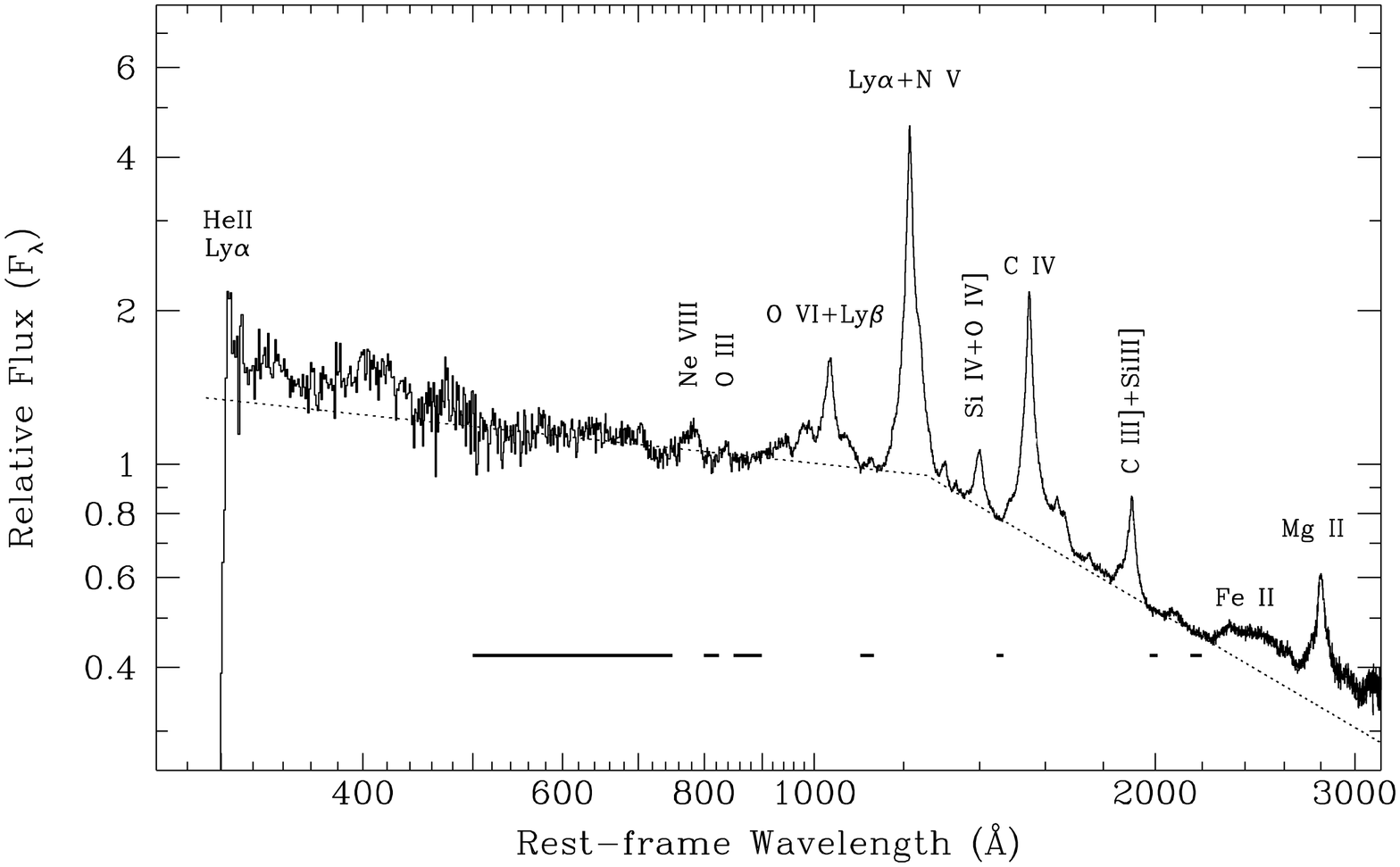,angle=-90,width=14cm}}
\caption{Overall mean composite QSO spectrum in 1 \AA\ bins with some prominent 
emission lines marked.  The dotted line shows the best fit broken power-law continuum, excluding 
the region below 500 \AA.  The lines at the bottom indicate the continuum windows used in the
fit.\label{fig:comp}}
\end{figure*}

The number of spectra contributing to this composite as a function
of wavelength is shown in Figure~\ref{fig:count}, and the S/N per 230 \kms\ 
is shown in Figure~\ref{fig:sn}.  The final overall composite is shown in Figure~\ref{fig:comp}
with some of the strong emission lines marked.  The S/N is computed per 
230 \kms\ since this roughly represents the resolution of the FOS when used with the 
high-resolution gratings.  The spikes in the S/N at the lowest wavelengths are real, and reflect
very high S/N data obtained to study the \ion{He}{2} Gunn-Peterson effect.

\begin{figurehere}
\centerline{\psfig{file=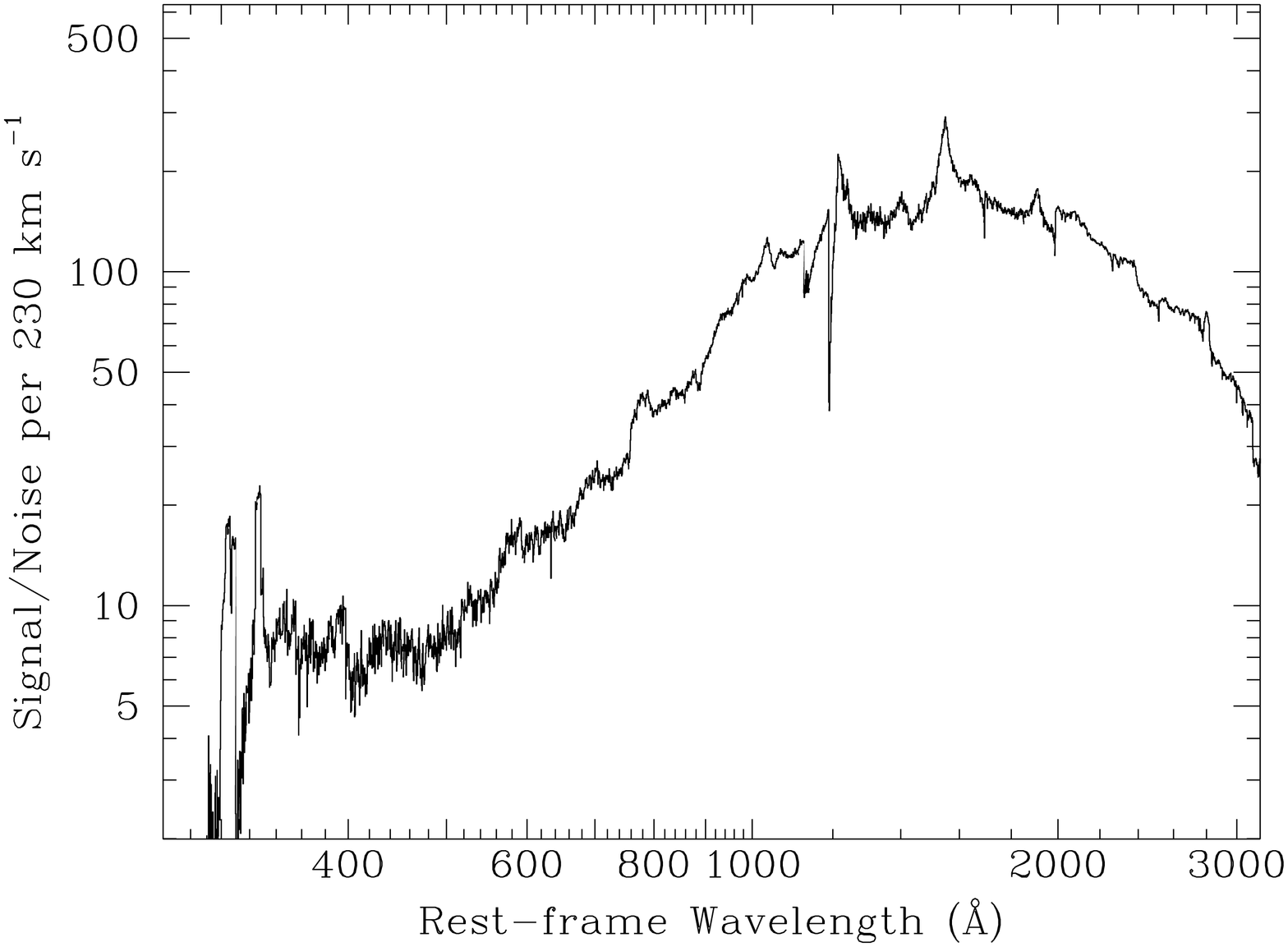, angle=-90, width=9cm}}
\caption{Signal-to-noise ratio in the overall composite per
bin of 230 \kms, approximately the resolution of the FOS with the high-resolution
gratings.\label{fig:sn}}
\end{figurehere}
\vspace{0.2cm}

The composite clearly shows the break in the power-law index
near \lya\ as reported by \zh, as do the subset composites that we discuss later.  
Therefore, to characterize the continuum, we fit the composite with a broken power law
using the IRAF task {\it specfit} \citep{kris94},
selecting wavelength windows free of emission lines for the fitting.  Specifically, the
wavelength windows we use are 350--750, 800--820, 850--900, 1095--1110, 1135--1150,
1450--1470, 1975--2010, and 2150--2200 \AA.   The exact location of the break is not
well determined due to the high density of emission lines in the region of the break.
For our fits the break wavelength varies from around 1200 to 1300 \AA, depending on the 
particular subset. 
Throughout this paper, we will refer to the power-law index longward of the break as 
$\alpha_{NUV}$ and that shortward of the break as $\alpha_{EUV}$, where both indices are 
defined by $f_\nu \propto \nu ^\alpha$.

The composite deviates from a broken power law with a slight hardening of 
the spectrum shortward of 500 \AA.  As can be seen in Figure~\ref{fig:count}, 
below 500 \AA\ fewer than ten spectra are contributing at any given wavelength.  The
fact that only a few spectra are contributing in this region is enough to raise doubt as 
to whether this portion of the composite is representative of all QSOs.  Equally
notable, however, is the significant flattening in the distribution of the number of spectra
with wavelength below 500 \AA.  Above 500 \AA, the shape of the distribution of the
number of spectra with wavelength is roughly what one would expect from a random sampling
of spectra of QSOs at various redshifts.  The flattening below 500 \AA\ is indicative of
the strong bias towards observing those few QSOs that are bright enough in this region
to obtain spectra with the HST, objects with preferentially hard UV spectra.  We
believe that the composite shape below 500 \AA\ is likely not representative of the
overall population of QSOs.  Therefore, in addition to the fit with the windows stated above,
we also produce a fit excluding the region below 500 \AA.

There are several possible choices as to the error
array to use in computing $\chi ^2$ for the fitting process.  We considered using
the propagated error array of the composite, reflected in Figure~\ref{fig:sn}. 
However, this has the undesirable result of giving too little weight to the shorter 
wavelengths ($\lesssim$ 800 \AA), both because many fewer objects contribute at these 
wavelengths and because the contributing spectra are generally of fainter, high-redshift
objects and therefore have inherently lower S/N.
We attempted using equal
weighting for all pixels, but this only reverses the problem; the fits are driven
strongly by the most extreme wavelengths.  The RMS deviation of the spectral fluxes is another
possible choice, but due to the fairly small number of objects contributing at the shortest 
wavelengths, this turns out to produce unreliable results.  The main problem is that $\chi ^2$
is often dominated by regions where there are only a few spectra contributing that happen
to have similar spectral shapes, resulting in a low value for the RMS deviation.
Our solution is to use an error 
array that consists of the flux of the composite, smoothed by 10 \AA\ to eliminate 
pixel-to-pixel noise, and divided by the square root of the number of contributing spectra.
The result is to give a weighting to each pixel in the fit that scales linearly
as the number of contributing spectra.  A different way of viewing this solution is that
the resulting errors are proportional to what would result from using the propagated
error array if the S/N of all the individual spectra were equal, thus eliminating the
bias towards high S/N spectra.  We find that with this
method we consistently obtain fits that characterize the spectrum well.  The best-fit indices, 
as well as some information about the sample, are shown as the first line in 
Table~\ref{ta:index}.  The quantity $\langle z \rangle$ is the mean redshift of the objects in
the sample, and $\langle L \rangle$ is the mean value of $\log \lambda L_{\lambda}$ at 1100 \AA\ 
for the sample objects in units of ${\rm erg \ s^{-1}}$, calculated as described in \S\ref{sec:lum}.
The best fit excluding the region below 500 \AA\ is shown as
the dotted line in Figure~\ref{fig:comp}, and the continuum windows used in the fit
are shown below the composite.

\begin{table*}
\caption{Fitted Power-Law Indices\label{ta:index}}
\centerline{
\begin{tabular}{lcccccccc} \hline \hline
&
Redshift	&	Number	&
Number	&
& & &
$>350$ \AA 	& $>500$ \AA\\
Radio		&
Range		&	spectra	&
objects		&
$\langle z \rangle$		&	$\langle L \rangle$	&
$\alpha_{NUV}$ &
$\alpha_{EUV}$ &		$\alpha_{EUV}$ \\
\hline
All	& All	& 332	& 184	& 1.17	& 46.16	&
$-0.69\pm 0.06$	& $-1.71\pm 0.13$	& $-1.76\pm 0.12$\\ 
RL	& All	& 205	& 107	& 1.00	& 46.01 &
$-0.67\pm 0.08$		& $-1.89\pm 0.15$		& $-1.96 \pm 0.12$\\
RQ	& All	& 127	& 77	& 1.42	& 46.38	&
$-0.72\pm 0.09$		& $-1.53\pm 0.16$		& $-1.57\pm 0.17$\\
RL	& $z<1.5$ & 190	& 94	& 0.84	& 45.91 &
$-0.67\pm 0.08$		& \nodata			& $-2.00\pm 0.09$\\
RQ	& $z<1.5$ & 79	& 45	& 0.84	& 46.03 &
$-0.72\pm 0.08$		& \nodata		&	$-1.70\pm 0.12$\\
\hline
Z97 \\
All	& All	& 184 	& 101 	& 0.93	& \nodata &
$-0.99$	& $-1.96$ & $-2.02^*$\\
RL	& All	& 110	& 60	& 0.87	& \nodata &
$-1.02$	& $-2.16$ & $-2.45^*$\\
RQ	& All	& 74	& 41	& 0.95	& \nodata &
$-0.86$ & $-1.77$ & $-1.83^*$\\ \hline
\\ 
\multicolumn{9}{l}{$^*$\zh\ obtained these indices by fitting above 600 \AA.}\\
\end{tabular}}
\end{table*}

\subsection{Uncertainties\label{sec:error}}

Because of the high S/N of the composite, the formal errors on the measured power-law indices
from the fits are negligibly small.  The uncertainties are thus dominated by the uncertainty
associated with producing a composite from our particular sample, as well as the flux corrections
that we apply.  We now discuss a few of these sources of uncertainty in detail.

The dominant uncertainty is that associated with our particular set of sample spectra, since the
scatter in spectral shapes for individual objects is quite large (see \S\ref{sec:indiv}).
To estimate the resulting uncertainty on the parameterization of the continuum, we implement the
bootstrap resampling technique.  Given the 332 spectra in the complete sample, we create 1000 random
samplings of these 332 spectra with replacement, fitting each one
as we did the true composite.  The results are distributions in the measured values of 
the power-law indices, and we calculate the $1 \sigma$ errors in the indices from these
distributions.

To investigate the degree to which our results depend on the Galactic extinction correction,
we repeated the analysis both with no extinction and with $E(\bv)$ doubled for each object.
We find that the EUV continuum is insensitive to these changes, as shown in 
Figure~\ref{fig:ext}.  The reason
for this at first surprising result is that since there are very few FOS G130H spectra 
in the sample, the EUV
portion of the composite is produced dominantly by G190H spectra.  The G190H bandpass contains
the 2200 \AA\ bump in the Galactic extinction curve, and as a result, the shape of the 
extinction curve is not monotonic.
The extinction correction for a G190H spectrum actually softens roughly as much of the 
continuum as it hardens, and as a result, the net change to the shape of the continuum is
very small.

A larger uncertainty is associated with the Lyman valley correction.  Many parameters
go into the correction, including those describing the shape of the \ion{H}{1} column density
distribution function ($\beta (N)$ in Equation~\ref{eq:valley}), the minimum and maximum column
densities,
the redshift evolution parameter ($\gamma$ in Equation~\ref{eq:valley}), the 
proportionality constant in Equation~\ref{eq:valley}, and the mean $b$ parameter of the
Lyman forest absorbers.  Of these parameters, those with uncertainties that could most affect our
results are the high-end cutoff column density and the redshift evolution parameterization.

\begin{figurehere}
\centerline{\psfig{file=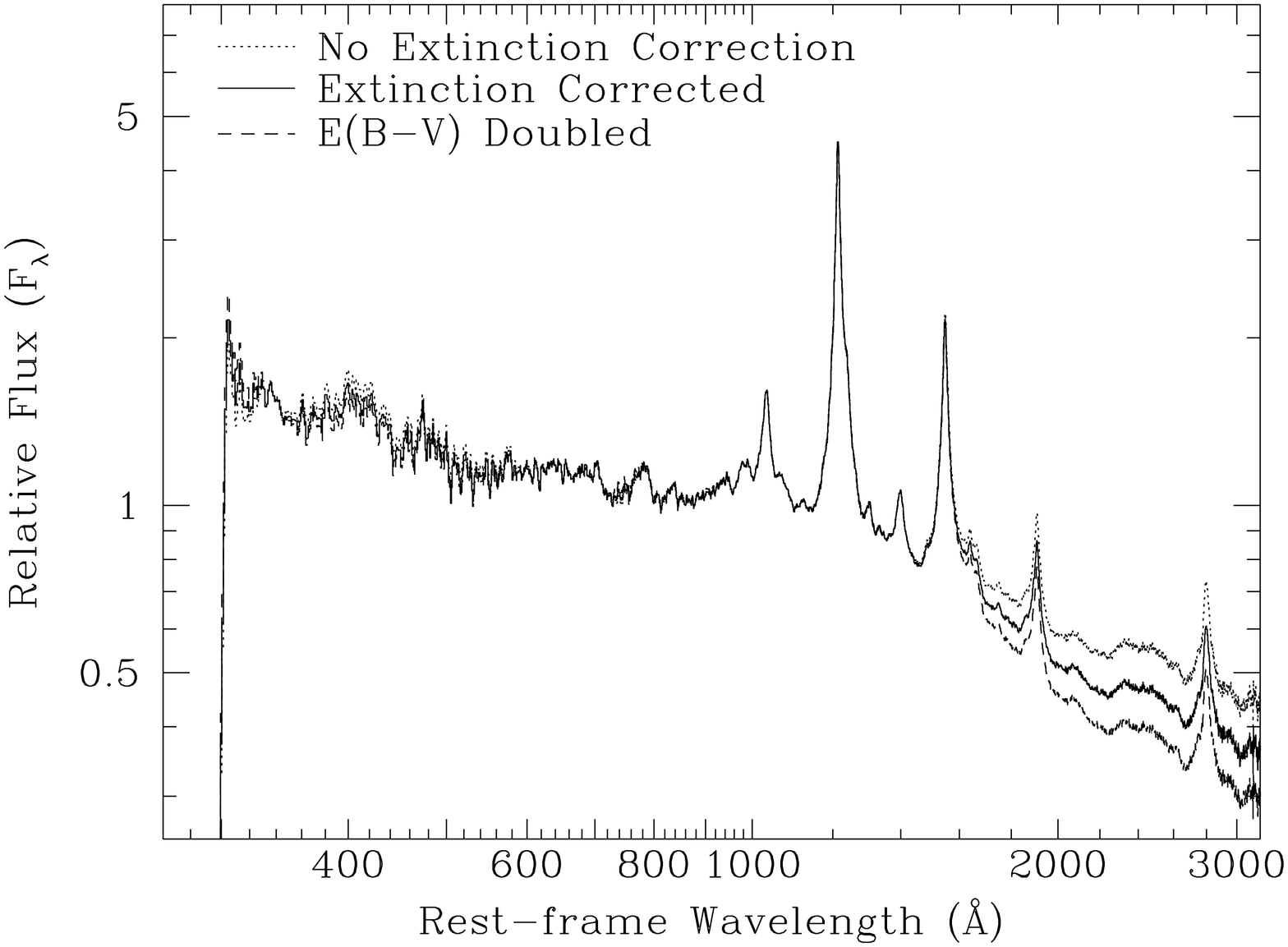,angle=-90,width=9cm}}
\caption{Overall composite (solid line), plotted along with the composite resulting from
applying no extinction correction (dotted line) and that resulting from doubling all values of 
$E(\bv)$ (dashed line).  The composites are virtually indistinguishable in the EUV.\label{fig:ext}}
\end{figurehere}
\vspace{0.2cm}

As we discussed in \S\ref{sec:reduce}, we correct statistically for all Lyman forest absorbers
with $N(HI) < 10^{16.7}$ \percm, since we have corrected for larger column density absorbers
individually.  However, we are not certain of being complete down to $10^{16.7}$ \percm,
and it could be that in fact a different limit is appropriate.  We tried increasing and decreasing
the limit by 50\%, 
thus encompassing the range $10^{16.5}$--$10^{16.9}$ \percm, and found that $\alpha_{EUV}$
in the total composite measured above 500 \AA\ varied from $-1.79$ to $-1.72$.  The 
error associated with the 
maximum column density is therefore at most of the order $\pm 0.04$, so we take this as a liberal
estimate of this error.

More poorly determined than the shape of the column density distribution function is
its evolution with redshift, usually parameterized by a power-law index $\gamma$
as in Equation~\ref{eq:valley}, or more generally $\gamma (z)$.  Low redshift studies with
HST have shown that $\gamma$ is small for $z < 1.5$, indicating very little evolution
\citep{bbbh+96,ipmw96}.  However, since the number of systems at low redshift is so much
smaller than for $z \gtrsim 2$, we concern ourselves only with the evolution at higher redshift.
For $z > 2$, accessible with ground-based spectroscopy, a large range in $\gamma$ has been
found, generally falling in the range 1.8--3.0, depending on the data and the details of the
analysis (\citealt{zulu93}; \citealt{bech94}; \citealt*{cec97}; \citealt{khcs97}).  
We chose $\gamma = 2.46$ from 
\citet{prs93} because it comfortably falls within the range of measured values.  To determine
the effect of $\gamma$ on the result, we also tried $\gamma = 2.1$ and $\gamma = 2.8$,
which resulted in $\alpha_{EUV}$ above 500 \AA\ in the total composite of $-1.73$ and 
$-1.78$, respectively.  We therefore estimate the error at around $\pm 0.03$.

The errors in $\alpha_{EUV}$ listed in Table~\ref{ta:index} include the bootstrap errors and
the errors associated with the Lyman valley correction as discussed above.  The errors in
$\alpha_{NUV}$ reflect only the bootstrap uncertainty, as this portion of the continuum is 
unaffected by the Lyman valley.

One may also wonder whether any individual objects could have a significant influence
on the composite.  In fact, the well-studied QSO HE~2347-4342 has such an extremely hard
EUV continuum that it does have a pronounced effect on the composite.  Removing it from
the overall sample softens the EUV power-law index, as fit above 500 \AA, from $-1.75$
to $-1.82$.  This effect is already accounted for in the errors in Table~\ref{ta:index},
since it is precisely this source of uncertainty that we are measuring with the bootstrap
resampling technique.  However, the effect of HE~2347-4342 on the composite raises the 
question of 
whether or not this sample is representative of the whole QSO population.  We return to 
this point in \S\ref{sec:indiv}.

\subsection{Luminosities\label{sec:lum}}

We estimate luminosities for the objects for the purpose of characterizing the sample.
Since the continuum turns over in the region $\sim 1250$ \AA, from $\alpha > -1$ above the
break to $\alpha < -1$ below, the energy distribution peaks near the break.  We therefore
use $\lambda L_{\lambda}$ at 1100 \AA\ as our estimator of the luminosity.  Since not all
spectra include 1100 \AA, we use the composite as a template for estimating the
flux at this wavelength.  We calculate the ratio of each spectrum to the composite over the
same wavelength range, then assume that this ratio also indicates the ratio of the fluxes
at 1100 \AA.  Thus, we use the following:
\begin{equation}
L_{\lambda} (1100 \\ \mbox{\AA}) = F_c(1100 \\ \mbox{\AA}) \frac{\sum_i F_{s,i}}{\sum_i F_{c,i}} 4 \pi D_L^2,
\end{equation}
where $F_c$ is the flux in the composite, $F_s$ is the flux in the individual spectrum,
and the summation is over the wavelength coverage of the spectrum, excluding the emission line
windows defined previously.  For an $\Omega_{\Lambda} = 0$ universe, the luminosity distance 
$D_L$ is given by the following formula first derived by \citet{matt58}:
\begin{equation}
D_L (z) = \frac{2c}{H_0 \Omega_0^2} \{\Omega_0 z + (\Omega_0 - 2)[(\Omega_0 z + 1)^{1/2} - 1]\},
\end{equation}
where we adopt $\Omega_0 = 1$ and $H_0 = 60 \ {\rm km \ s^{-1} \ Mpc^{-1}}$.  This yields
a luminosity for each spectrum.  For objects for which there is just one spectrum, that 
luminosity is assigned to the object.  For those with more than one spectrum, 
if one of the spectra contains 1100 \AA, the luminosity obtained from that spectrum is 
assigned to the object, otherwise the luminosity from the spectrum with wavelength coverage
closest to 1100 \AA\ is used.  In Figure~\ref{fig:zlumdist} we plot the luminosities of all
the objects in our sample as a function of redshift.  The strong correlation of luminosity
with redshift characteristic of flux-limited samples is apparent here.  It is interesting to
note the distribution of radio-loud and radio-quiet objects with redshift.  At intermediate
redshifts, $0.6 < z < 1$, the sample is dominated by radio-loud QSOs, whereas for $z \gtrsim 1.5$,
the sample is mostly radio-quiet.  As a result, at longer wavelengths, the overall composite
is dominated by radio-loud data, while at shorter wavelengths (specifically, 
$\lambda \lesssim 900$ \AA) the composite contains more radio-quiet data than radio-loud.

\begin{figurehere}
\centerline{\psfig{file=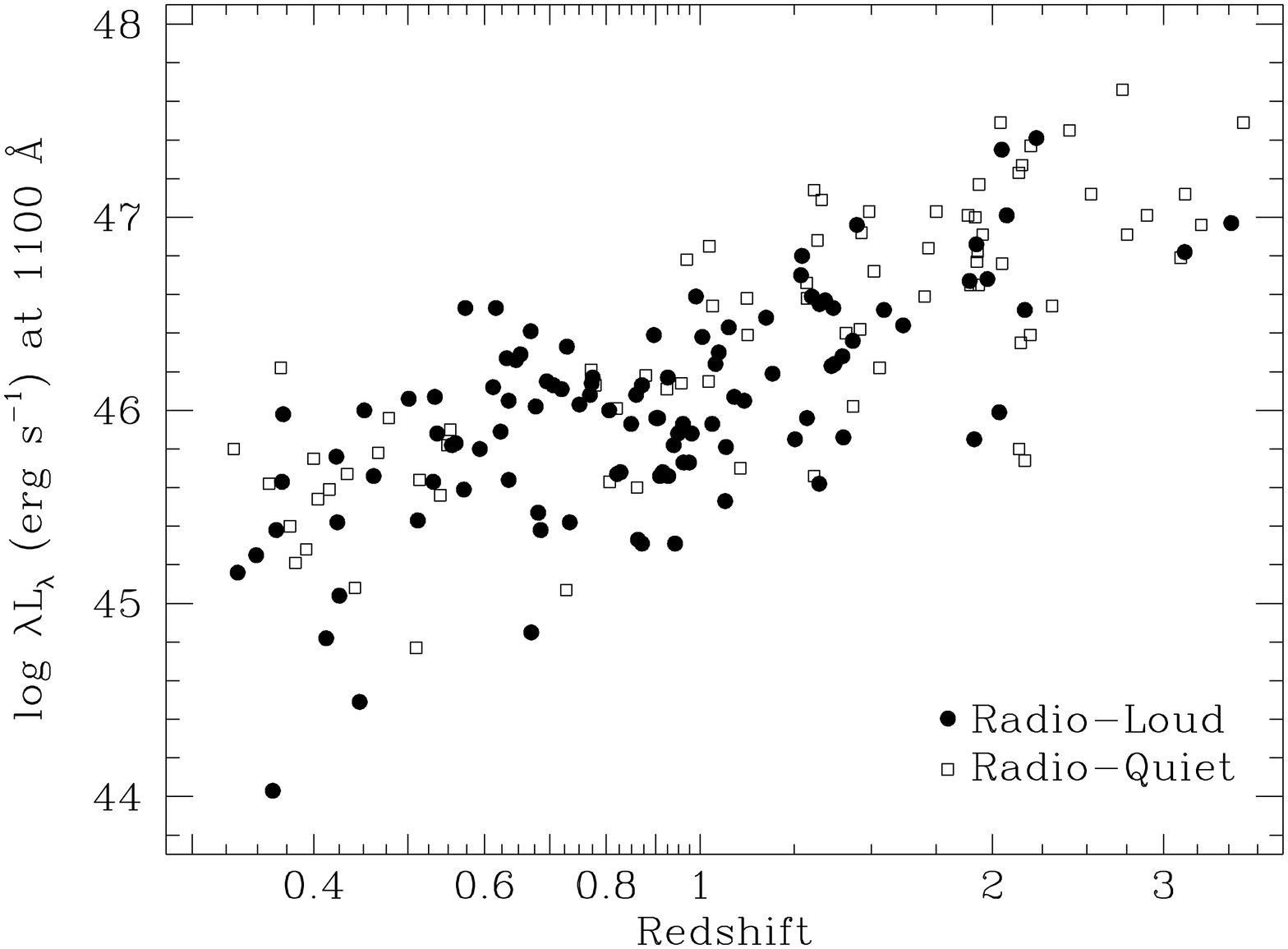,angle=-90,width=9cm}}
\caption{Distribution of QSOs in redshift and monochromatic luminosity at
1100 \AA.  Filled circles are radio-loud objects, open squares are 
radio-quiet.\label{fig:zlumdist}}
\end{figurehere}
\vspace{0.2cm}

\subsection{Radio properties\label{sec:radio}}

Given the large numbers of each at our disposal, we separate the sample into radio-loud
and radio-quiet subsamples to investigate the radio dependence.  The analysis of
each subset is done as it was for the full composite, creating the composites with a bootstrap
and fitting with a broken power law.  The bootstrap errors were also computed independently for 
each sample as for the full composite.  The full radio-loud and radio-quiet composites are
plotted in Figures~\ref{fig:rl} and \ref{fig:rq}, 
and the power-law indices are listed in Table~\ref{ta:index}.
The radio-loud composite has a significantly softer EUV continuum than the radio-quiet 
composite, although the NUV continua are virtually identical.
However, as is evident in Figure~\ref{fig:zlumdist}, the distribution of objects with redshift
is quite different for the two groups above $z \gtrsim 1.5$.  We therefore repeated the analysis
using only objects for which $z < 1.5$, for which the distributions in redshift and
luminosity are quite similar.  The results are also listed in Table~\ref{ta:index}.  The
difference between the two is still substantial although somewhat smaller for 
$z < 1.5$ than for the full composites, primarily due
to the exclusion of HE~2347-4342 from the radio-quiet sample.
Using the bootstrap resampling distributions used to compute the errors
on $\alpha_{EUV}$, we estimate that the radio-loud composite is softer than the radio-quiet
composite with 98.6\% confidence.

\begin{figurehere}
\centerline{\psfig{file=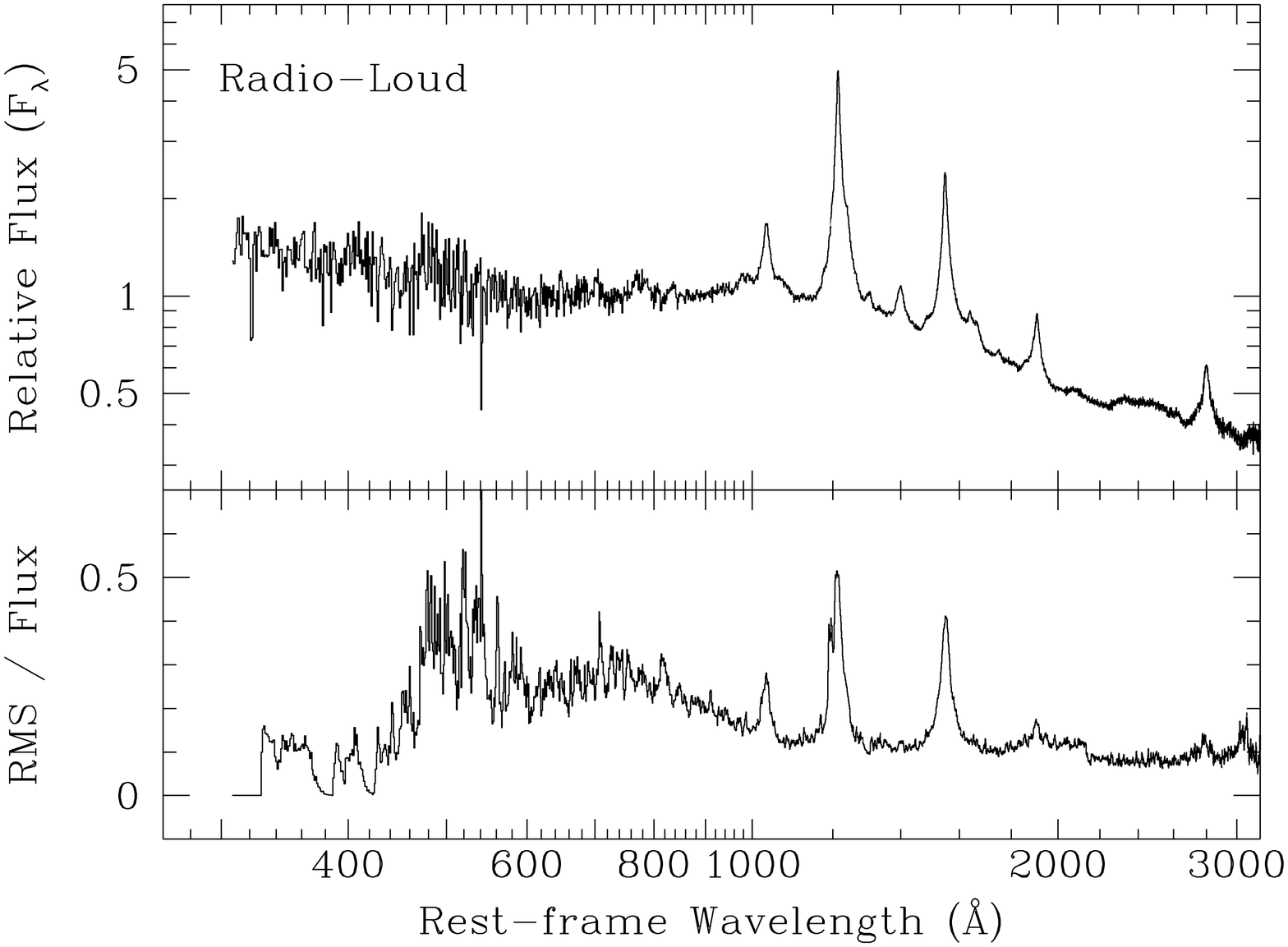,angle=-90,width=9cm}}
\caption{Radio-loud composite spectrum (above) and corresponding relative RMS deviation from the
mean for each 1 \AA\ pixel (below).  The RMS deviation has been smoothed by five pixels.
\label{fig:rl}}
\end{figurehere}
\vspace{0.2cm}

\begin{figurehere}
\centerline{\psfig{file=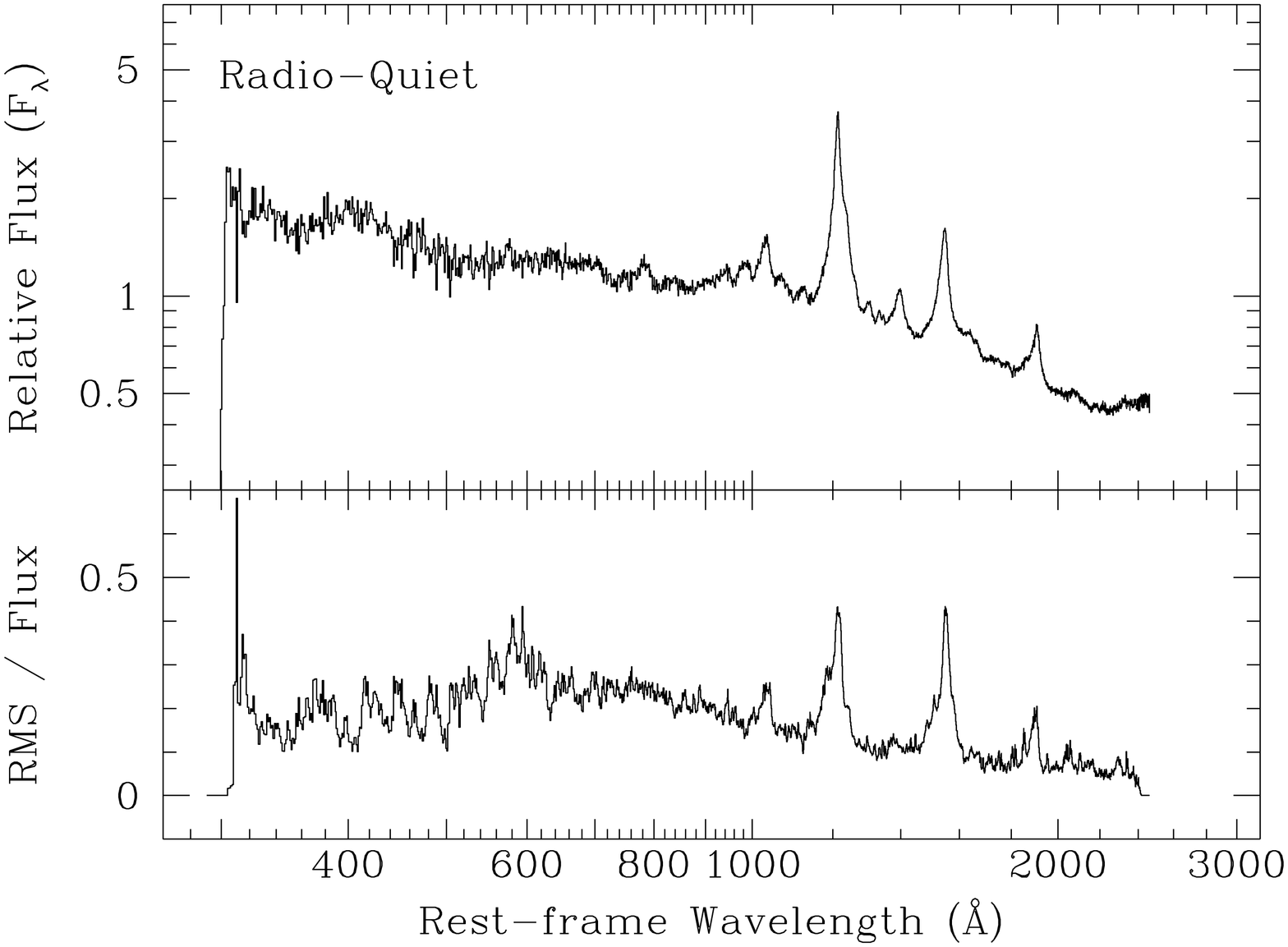,angle=-90,width=9cm}}
\caption{Same as Figure~\ref{fig:rl} for the radio-quiet composite.\label{fig:rq}}
\end{figurehere}
\vspace{0.2cm}

\subsection{Spectral Variations\label{sec:vary}}
In addition to the mean composites, we also plot in Figures~\ref{fig:rl} and \ref{fig:rq} the
RMS deviation of the individual
spectra about the mean in 1 \AA\ bins for the radio-loud and radio-quiet subsamples, respectively, 
smoothed by five pixels and normalized to the composite flux.  The RMS deviations in
the two groups are very similar, differing significantly only where the number of contributing 
objects is small.  The RMS spectra consist mostly of an RMS ``continuum'' caused by a 
combination of differences in continuum shapes and noise, both Poisson noise and that resulting
from intervening absorption.  The RMS continuum increases towards shorter wavelengths down
to $\sim 700$ \AA, where the Lyman valley correction peaks.  Shortward of this, there are large
fluctuations in the RMS deviation as the number of contributing objects becomes small.
Above the continuum, one can see the effects of the variation in the strong emission lines,
particularly \ion{O}{6}, \lya, and \ion{C}{4}.  By subtracting in quadrature the RMS continuum
from the emission line RMS values, we can estimate the RMS deviation in the strength of the
emission lines.  The features in Figures~\ref{fig:rl} and \ref{fig:rq} correspond to RMS
fluctuations in the peak emission-line flux in \ion{O}{6}, \lya, and \ion{C}{4} of 
$\sim 50$--$70$\%.

\subsection{Emission Lines\label{sec:emline}}

To aid in making likely identifications of the emission lines in our composites, we generate
a few simple broad-line region models using typical parameters with the 
photoionization code CLOUDY \citep[version 94.00;][]{ferl96}.  We use single-slab models 
of a broad-line region with solar abundances, densities of $10^8$ to $10^{10}\ {\rm cm^{-3}}$,
and total column densities from $10^{23}$ to $10^{25.5}$ \percm.  We assume the broad-line 
region is illuminated by a power-law continuum with index $\alpha_{\nu}$ from $-1.5$ to 
$-1.8$ and an ionization parameter ranging from $\log U = -2.0$ to 0.0.  We look for the 
strongest lines in the models, those with fluxes greater than $\sim 0.001$ times H$\beta$, and 
consider these to be the most plausible identifications.  For the plentiful transitions of 
\ion{Fe}{2}, \ion{Fe}{3}, \ion{Si}{2}, and \ion{Si}{3}, we use as a guide the identifications 
made in the SDSS composite \citep{vand+01} and in the narrow-line quasar 1~ZW~1 \citep{ljgb97}.

\begin{table*}
\caption{Emission Lines\label{ta:emline}}
\small{\centerline{
\begin{tabular}{lccccccc} \hline \hline
			&				&
\multicolumn{2}{c}{Whole Sample}			&
\multicolumn{2}{c}{Radio Quiet}				&
\multicolumn{2}{c}{Radio Loud}\\
 \cline{3-4} \cline{5-6} \cline{7-8}
			&	$\lambda_0$	&
			&	EW		&
			&	EW		&
			&	EW \\
Line		&	(\AA)		&
Flux		&	(\AA)		&
Flux		&	(\AA)		&
Flux		&	(\AA)\\ \hline
\ion{Ne}{8} + \ion{O}{4} & $\sim 780$ & $5.2\pm 0.2$ & 4.4 & 
$5.7 \pm 0.2$ & 4.3 & $4.7 \pm 0.3$ & 4.1 \\
\ion{O}{3}	& 835 & $1.1 \pm 0.1$ & 1.0 &
$1.5 \pm 0.2$ & 1.2 & $1.2 \pm 0.1$ & 1.0 \\
\ion{H}{1} Ly series + \ion{S}{6} & $\sim 930$ & $5.1 \pm 0.2$ & 4.3 &
$8.4 \pm 0.3$ & 6.6 & $3.1 \pm 0.3$ & 2.7 \\
\ion{C}{3} + \ion{N}{3} & $\sim 980$ & $11.3 \pm 0.2$ & 9.7 &
$11.7 \pm 0.2$ & 9.4 & $8.1 \pm 0.6$ & 7.1 \\
\ion{O}{6} + \lyb	& $\sim 1030$ & $18.0 \pm 0.3$ & 15.6 &
$18.1 \pm 0.9$ & 14.6 & $19.1 \pm 2.6$ & 16.9 \\
\ion{N}{2} + \ion{He}{2} + ? & $\sim 1065$ & $5.2 \pm 0.2$ & 4.6 &
$5.5 \pm 0.2$ & 4.5 & $5.6 \pm 1.1$ & 5.0 \\
\ion{Fe}{3} & 1123 & $0.65 \pm 0.06$ & 0.57 &
$2.2 \pm 0.2$ & 1.8 & $0.28 \pm 0.04$ & 0.25 \\
\ion{C}{3}$^{*}$ & 1176 & $0.40 \pm 0.05$ & 0.36 &
$0.43 \pm 0.05$ & 0.36 & $0.44 \pm 0.03$ & 0.39 \\
\ion{Si}{2} & $\sim 1195$ & $1.02 \pm 0.08$ & 0.93 &
$0.47 \pm 0.09$ & 0.40 & $1.5 \pm 0.1$ & 1.4 \\
\lya	& 1216 & $100.0 \pm 0.8$ & 91.8 &
$100.0 \pm 0.8$	& 86.3 & $100.0 \pm 0.7$ & 91.2 \\
\ion{N}{5} & 1240 & $20.0 \pm 0.5$ & 18.5 &
$17.5 \pm 0.6$ & 15.4 & $22.0 \pm 0.4$ & 20.2 \\
\ion{Si}{2} & 1263 & $0.32 \pm 0.02$ & 0.30 &
$0.41 \pm 0.03$ & 0.37 & $0.27 \pm 0.02$ & 0.25 \\
\ion{O}{1} + \ion{Si}{2} & $\sim 1305$ & $2.03 \pm 0.07$ & 1.94 &
$2.5 \pm 0.1$ & 2.3 & $1.9 \pm 0.1$ & 1.7 \\
\ion{C}{2} & 1335 & $0.61 \pm 0.05$ & 0.60 &
$1.0 \pm 0.4$ & 0.9 & $0.35 \pm 0.03$ & 0.34 \\
? & 1347 & $0.15 \pm 0.02$ & 0.15 &
$0.3 \pm 0.1$ & 0.3 & $0.07 \pm 0.02$ & 0.07 \\
\ion{Si}{4} + \ion{O}{4}] & $\sim 1400$ & $9.4 \pm 0.2$ & 9.9 &
$11.9 \pm 1.3$ & 12.2 & $8.6 \pm 0.3$ & 9.0 \\
\ion{N}{4}] & 1486 & $2.4 \pm 0.1$ & 2.7 &
$0.6 \pm 0.1$ & 0.6 & $2.8 \pm 0.1$ & 3.2 \\
\ion{C}{4} & 1549 & $48.0 \pm 0.4$ & 58.0 &
$38.0 \pm 4.4$ & 44.7 & $52.0 \pm 0.2$ & 62.9 \\
broad feature & $\sim 1600$ & $26.7 \pm 0.5$ & 34.6 &
$25.9 \pm 0.9$ & 31.9 & $30.2 \pm 0.4$ & 38.5 \\
\ion{He}{2} & 1640 & $1.11 \pm 0.05$ & 1.45 &
$1.0 \pm 0.5$ & 1.3 & $1.19 \pm 0.05$ & 1.57 \\
\ion{O}{3}] & 1664 & $1.79 \pm 0.03$ & 2.39 &
$0.7 \pm 0.6$ & 1.0 & $2.25 \pm 0.03$ & 3.01 \\
\ion{N}{3}] & 1750 & $0.44 \pm 0.02$ & 0.58 &
$0.21 \pm 0.02$ & 0.26 & $0.50 \pm 0.02$ & 0.67 \\
\ion{Al}{3} & 1857 & $1.49 \pm 0.05$ & 2.19 &
$2.4 \pm 0.1$ & 3.3 & $1.20 \pm 0.05$ & 1.77 \\
\ion{C}{3}] + \ion{Si}{3}] & $\sim 1900$ & $12.9 \pm 0.2$ & 19.7 &
$12.8 \pm 0.4$ & 18.5 & $13.2 \pm 0.2$ & 20.3 \\
\ion{Fe}{2} & $\sim 2185$ & $1.7 \pm 0.1$ & 3.2 &
$0.60 \pm 0.04$ & 1.1 & $1.8 \pm 0.1$ & 3.4\\
\ion{Mg}{2} & 2799 & $22.6 \pm 0.2$ & 51.7 &
\nodata & \nodata & $22.3 \pm 0.3$ & 51.7 \\ \hline
\end{tabular}}}
\end{table*}

The composite is shown in Figures~\ref{fig:euvemline} and \ref{fig:nuvemline} with all
of the identified emission lines between 600 and 2000 \AA\ labeled.  We now make some
brief comments on a few features:
\begin{enumerate}
\item{There is an odd feature at $\sim 700$ \AA, a slight change in flux that looks like it could
be partly due to an emission line, which we tentatively identify as \ion{O}{3} \lam 703.
We have tried changing the details of the composite construction
process, including the direction of the bootstrapping, the weighting of the individual
spectra, and the emission line masking windows, but the feature always appears.  It therefore seems 
unlikely that this is due to a problem with our procedure, such as improper normalization of the
spectra.  Also, there are more than 40 spectra contributing at this wavelength, so the feature
should not be due to any single object, and we have explicitly checked that this is the case.
Thus, it seems that this feature is representative of the sample spectra.}
\item{We identify the
feature at $\sim 774$ \AA\ as primarily \ion{Ne}{8} as has been previously reported
(\citealt*{hzt95}; \citealt{hcsb+98}).  However, the models predict some \ion{O}{4}
\lam 789 to be present, and the appearance of the \ion{Ne}{8} feature as being skewed and 
shifted to the red suggests that \ion{O}{4} may contribute.}
\item{A broad feature from $\sim 900$ to $\sim 950$ \AA\ appears in the composite that
we identify as collected emission of many \ion{H}{1} Lyman lines of order Ly$\delta$
and higher, possibly blended with \ion{S}{6} \lamlam 933, 945.
The \ion{He}{2} Balmer series could contribute,  but due to the lack of evidence 
for emission from \ion{He}{2} Ba7 at 959 \AA, we conclude that it is unimportant.}
\item{Redward of \ion{O}{6} at around 1065 \AA\ appears a broad feature that can only
partially be explained by \ion{N}{2} and \ion{He}{2} at 1085 \AA.  \zh\ tentatively identified
this feature as \ion{Ar}{1} \lam 1067.  We find no definitive identification, though it
could be due to an \ion{Fe}{2} blend visible in the theoretical \ion{Fe}{2} simulations
of \citet{vvkf+99}.  Another possible identification is 
\ion{S}{4} \lamlam 1063, 1073 \citep{zkwb+01}.}
\item{There is a weak emission feature at $\sim 1347$ \AA\ that we cannot identify.  A similar
feature was observed by \citet{ljgb97}.}
\end{enumerate}

To measure the strength of the emission features we fit them with Gaussian components
using {\it specfit}.  The resulting emission-line fluxes and
equivalent widths are listed in Table~\ref{ta:emline}, as well as the results for the
radio-loud and radio-quiet composites.  We do not quote the errors in equivalent width because the
relative errors are the same as for the fluxes.  The fluxes are normalized to a flux of 
100 for \lya.
In producing the fits, we use the simplest model possible while still providing a good fit 
to the data.  

The spectrum is fit in five pieces:
730--1150 \AA, 1140--1323 \AA, 1318--1705 \AA, 1700--2215 \AA, and 2100--3020 \AA.
Each section is fit with a power-law continuum plus Gaussian components.  \ion{Mg}{2},
\ion{C}{3}], \ion{Si}{4} + \ion{O}{4}], \ion{C}{4}, \ion{N}{5}, and \ion{O}{6}
are all fit with
two Gaussian components, a broad one and a narrower one, with the peak wavelengths of
the two components fixed to each other.  \lya\ requires an additional narrow
component for a good fit.  In fitting the \lya\ / \ion{N}{5} feature, the wavelength ratio
of the \lya\ to the \ion{N}{5} components is fixed at its laboratory value, and the widths of
the \ion{N}{5} components are bound to the widths of the broader two \lya\ components.
The wavelength ratios of \lyb\ to \ion{O}{6} and \ion{Si}{3}] to \ion{C}{3} are fixed in the
same way.
Because the \ion{C}{4} profile is so skewed, especially in the radio-quiet composite,
it is necessary to fit \ion{C}{4} with a skewed Gaussian where the skew parameters of each of the 
components are bound together.  All of the remaining features are fit with a single
Gaussian component, including \ion{Ne}{8} + \ion{O}{4},  the broad feature at
$\sim 930$ \AA, \ion{C}{3} + \ion{N}{3}, \ion{Si}{4} + \ion{O}{4}], and the broad feature 
longward of \ion{O}{6} including \ion{N}{2} and \ion{He}{2}.  
\ion{Ne}{8} + \ion{O}{4} and the \ion{H}{1} Lyman feature were also skewed
to provide a good fit.  The addition of a very broad feature (FWHM $\sim 25000$ \kms)
at $\sim 1600$ \AA, perhaps due to \ion{Fe}{2} emission or a very broad component (VBC) of 
\ion{C}{4}, is necessary to provide a good fit to the \ion{C}{4} blend.  This component
comprises a rather large amount of flux, more than half that of \ion{C}{4}.  The resulting
decomposition of the \ion{C}{4} blend is quite similar to that performed by \citet{lbjs+94}
on similar data.
We quote summed values for some blends where the deblending is
highly uncertain.  For some of the weaker lines, the values for the overall composite do not
fall in between those for the radio-loud and radio-quiet composites, a result of some
uncertainty in the modeling not reflected in the quoted errors.

In addition to comparing line strengths of the radio-loud and radio-quiet composites, it is
instructive to compare the spectra directly.  We have plotted the radio-loud and 
radio-quiet composites together in Figures~\ref{fig:euvemline} and \ref{fig:nuvemline}.
The ratios of the two are shown in the bottom panels of the figures.

There are some noteworthy differences between the different radio composites.  \citet{fhi93} 
found that \lya\ and \ion{C}{4} have both larger equivalent widths and stronger narrower 
peaks in radio-loud objects.  We verify these results, except that the difference in \lya\
equivalent width is quite moderate, only 6\%.  However, we note that the sample of 
radio-loud QSOs for which \citet{fhi93} measured \lya\ is quite small (only seven objects).
The enhancement of the equivalent width of \ion{C}{4}, 41\% larger in radio-loud QSOs than
radio-quiet QSOs, agrees well with their results.  Equivalent width enhancements for
\ion{N}{5} (31\%) and \ion{O}{6} (16\%) are also seen in our data.
As can be seen in Figure~\ref{fig:euvemline},
the tendency for a stronger narrow component in radio-loud objects
may also apply to \ion{O}{6}, although much less than for \ion{C}{4} and \lya.

Also notable is the difference in the distribution of the emission in velocity space.
The profiles of \ion{O}{6}, \lya, \ion{Si}{4} + \ion{O}{4}], \ion{C}{4}, and \ion{C}{3}]
all show a tendency for more emission to the red
and less to the blue in radio-loud QSOs as opposed to radio-quiet.  For \ion{C}{3}], this
could be due to a difference in the neighboring \ion{Si}{3}] and \ion{Al}{3} emission.  Also, 
there are only twelve spectra contributing to the \ion{C}{3}] profile in the radio-quiet sample, so this
may not be a general property.
For \ion{O}{6}, \lya, and \ion{Si}{4} + \ion{O}{4}], this can be mostly explained by a systematic
blueshift of the entire line emission in radio-quiet objects,
an effect commonly observed in \ion{C}{4} \citep[e.g.,][]{msdc+96, szmd00}.  However,
in our data, the difference in \ion{C}{4} emission between radio-loud and radio-quiet
objects appears to be due largely to a blue asymmetry in the radio-quiet profile shape,
an effect which has also been reported \citep{wbfs+93}.

There are other minor differences as well.  The broad feature at
$\sim 930$ \AA\ is quite strong in the radio-quiet composite, but more than a factor of two weaker 
in the radio-loud composite.  The \ion{Fe}{3} \lam 1123 feature is also much stronger in the
radio-quiet composite, by roughly a factor of seven.

\section{INDIVIDUAL OBJECTS\label{sec:indiv}}

To explore in another way the difference in $\alpha_{EUV}$ between radio-loud and radio-quiet,
as well as to search for correlations of $\alpha_{EUV}$ with redshift and luminosity, we
fit $\alpha_{EUV}$  for all 
individual objects for which it is possible to do so using the same continuum windows as for
the composite.  In order to have a sufficient baseline for measuring $\alpha_{EUV}$, we
demand that to fit an object it must either have either (1) data above 1100 \AA\ and data below
900 \AA, or (2) continuous data from 700 \AA\ to 900 \AA.  For a few objects, it was 
necessary to alter slightly the continuum windows, so as to avoid absorption features,
in order to obtain a good fit.  A few others had to be thrown out because absorption was
too prevalent or the S/N was too low to obtain a good fit.  For performing these fits
for objects for which there are multiple spectra with continuous wavelength coverage,
the individual spectra are combined into a single spectrum.

\begin{table*}
\caption{Individual Object Fits\label{ta:ind}}
\centerline{
\begin{tabular}{lcccccccccc} \hline \hline
			&		&			&	&
			&
\multicolumn{3}{c}{$\alpha_{EUV}$ vs.\ $\log z$} &
\multicolumn{3}{c}{$\alpha_{EUV}$ vs.\ $\log L$}\\
\cline{6-8} \cline{9-11}
&& Mean & Median &&&&&&&\\
Group			& N	& $\alpha_{EUV}$	& $\alpha_{EUV}$ &
RMS	& 
$r$		& Signif.\	& 	Slope	&
$r$		& Signif.\	&	Slope\\ \hline
RQ & 39 & $-1.61\pm 0.14$ & $-1.59 \pm 0.06$ & 0.86 & 
0.118 & 0.527 & $0.09 \pm 0.56$ & -0.004 & 0.017 & $-0.03 \pm 0.27$\\
RL & 40 & $-1.95\pm 0.11$ & $-1.95 \pm 0.12$ & 0.66 & 
-0.119 & 0.534 & $-0.33 \pm 0.66$ & 0.271 & 0.909 & $0.63 \pm 0.27$\\ \hline
\end{tabular}}
\end{table*}

The histograms of $\alpha_{EUV}$ for the 40 radio-loud and 39 radio-quiet objects that 
could be fit are
shown in Figure~\ref{fig:alphahist}.  The mean, median, and standard deviation of these 
distributions are listed in Table~\ref{ta:ind}.  We estimate the errors on the mean and median
for each distribution by again using the bootstrap resampling method, except that for the median
we use the half-sample method, which is a better estimator for the error on the median
\citep{babu92}.  The means and medians of the distributions
agree well with the indices fit to the composites, which argues that the composites are
indeed representative of their constituent spectra.  A two-population Kolmogorov-Smirnov test
shows that the distribution of $\alpha_{EUV}$ for the radio-loud and radio-quiet objects
are different with 98.6\% confidence.

\begin{figurehere}
\centerline{\psfig{file=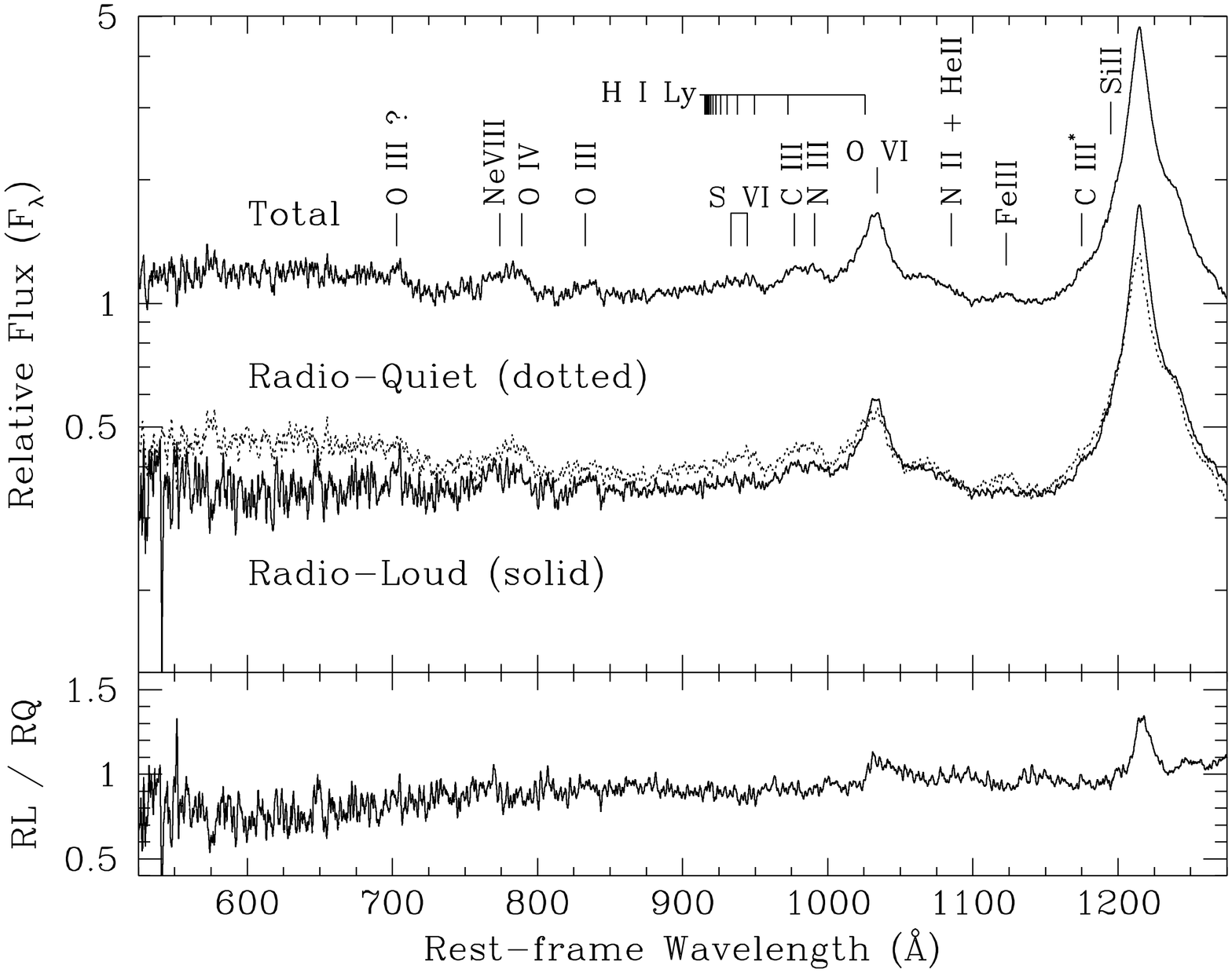,angle=-90,width=9cm}}
\caption{The top panel shows the overall composite below \lya\ with identifed
emission lines marked.  Below the
overall composite are the radio-loud composite (solid line) and 
the radio-quiet composite (dotted line), normalized to have the same
flux at $\sim 1100$ \AA.  The bottom panel shows the ratio of the radio-loud composite to
the radio-quiet composite.\label{fig:euvemline}}
\end{figurehere}
\vspace{0.2cm}

\begin{figurehere}
\centerline{\psfig{file=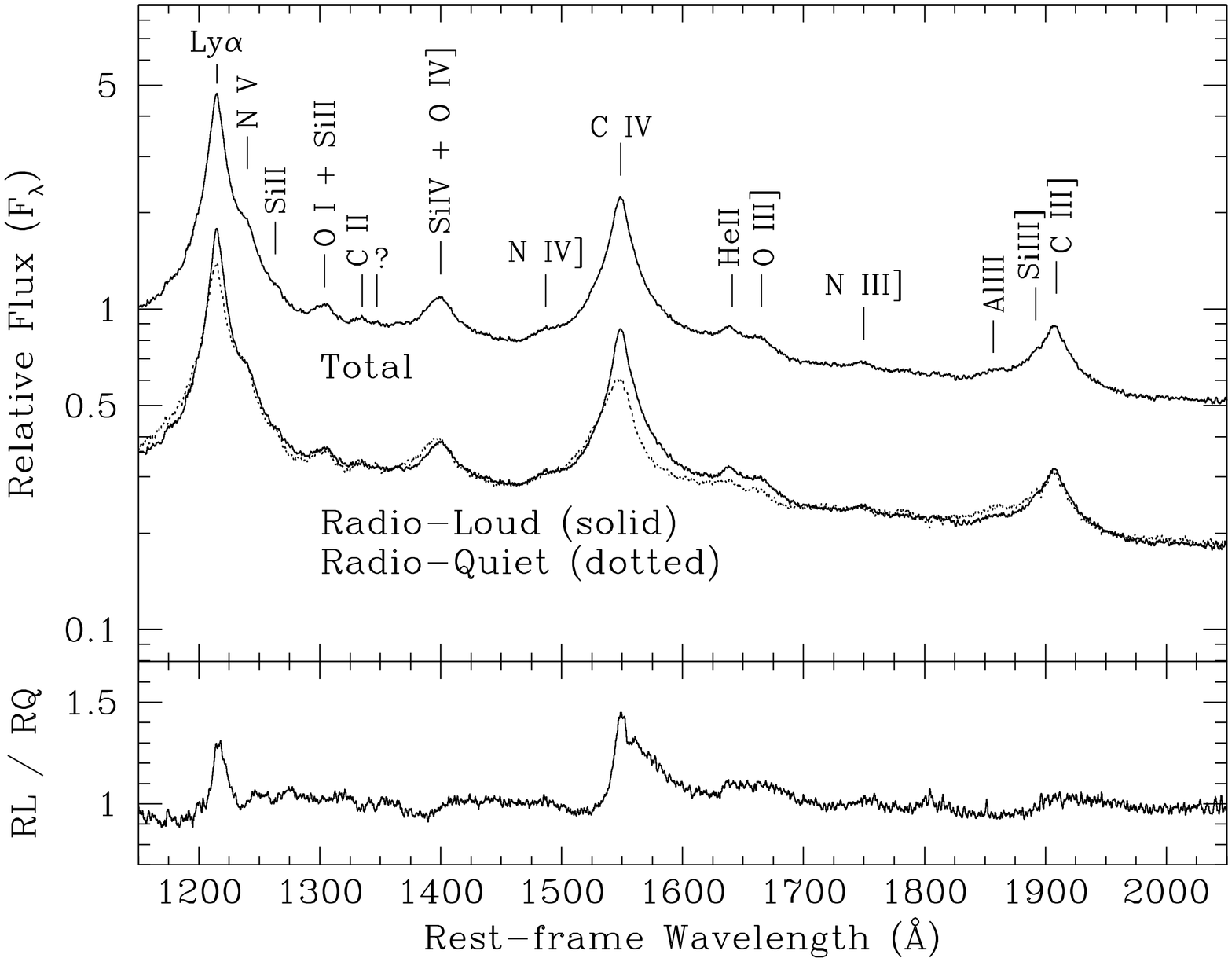,angle=-90,width=9cm}}
\caption{Same as Figure~\ref{fig:euvemline} for the spectral region from \lya\ to
2050 \AA.  The radio-loud and radio-quiet composites are normalized to have the same flux
at 1450 \AA.  Differences in the profiles of the strong broad lines are evident, 
particularly for \ion{C}{4}.\label{fig:nuvemline}}
\end{figurehere}
\vspace{0.2cm}

The objects are plotted in the redshift-$\alpha_{EUV}$ parameter space in 
Figure~\ref{fig:alphared},
and likewise they are plotted in the luminosity-$\alpha_{EUV}$ parameter space in
Figure~\ref{fig:alphalum}.  To test if there is any correlation between $\alpha_{EUV}$ and
either redshift or luminosity, we calculate the Spearman rank-order correlation coefficient $r$
and its significance for each relation, listed in Table~\ref{ta:ind}.  Only one relation,
$\alpha_{EUV}$ vs.\ luminosity for radio-loud objects, shows any evidence for a correlation,
and only marginally so, at the $\sim 90$\% confidence level.  
To place limits on the possible range of any variation of $\alpha_{EUV}$ with redshift or 
luminosity for comparison with the literature, we fit $\alpha_{EUV}$ vs.\ $\log z$ and 
$\alpha_{EUV}$ vs.\ $\log L$ using a standard
least-squares linear regression.  The best fit slopes and errors are also shown in 
Table~\ref{ta:ind}.  Consistent with the rank-order correlation test, all of the slopes are 
consistent with zero with the exception of $\alpha_{EUV}$ vs.\ $\log L$ for the radio-loud objects.

\begin{figurehere}
\centerline{\psfig{file=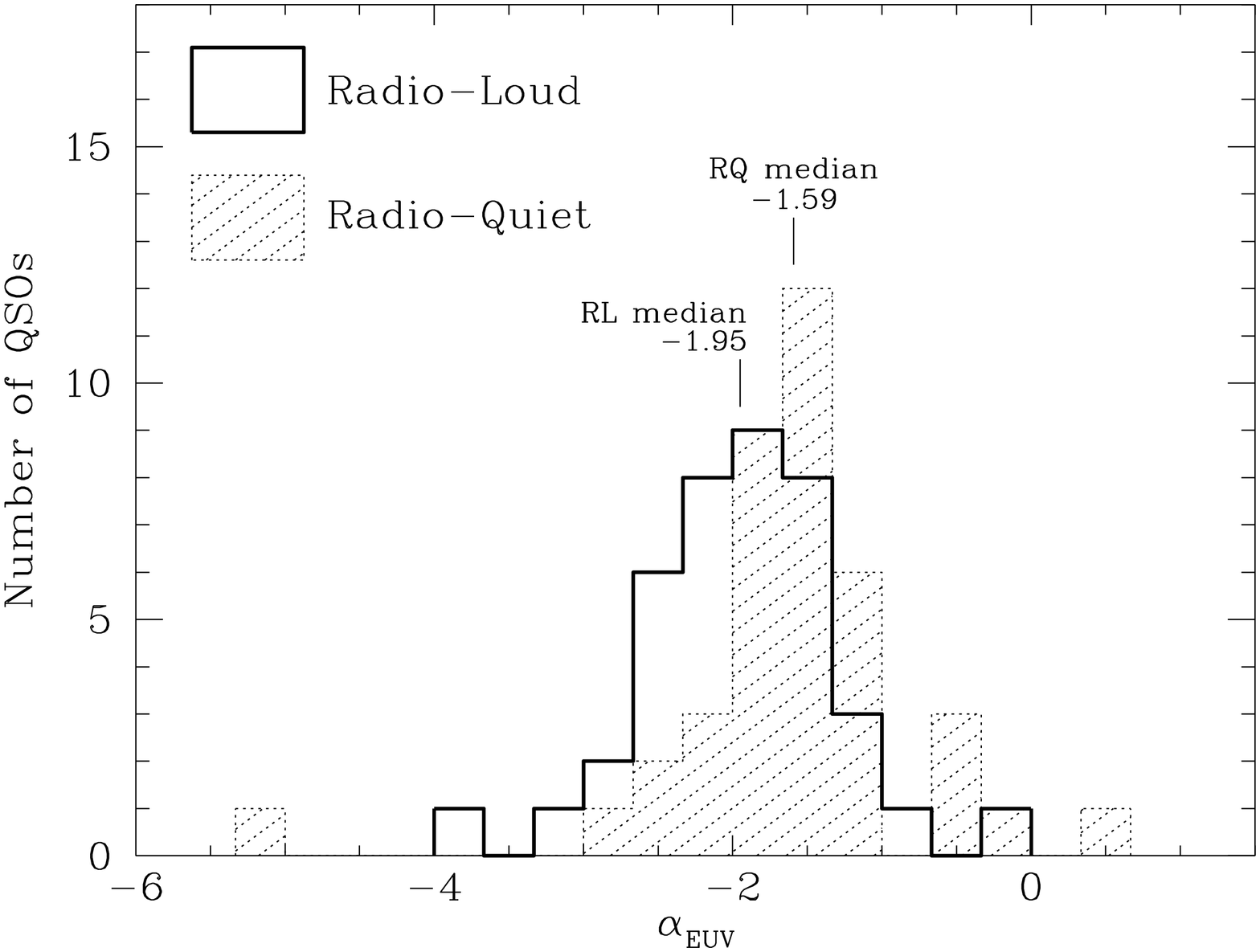,angle=-90,width=9cm}}
\caption{Histogram of EUV power-law indices in bins of 0.33 for 40 radio-loud
and 39 radio-quiet QSOs.  The means, medians, and RMS deviations of the distributions are
listed in Table~\ref{ta:ind}.\label{fig:alphahist}}
\end{figurehere}
\vspace{0.2cm}

In Figures~\ref{fig:alphared} and \ref{fig:alphalum} we have labeled three objects that are
noteworthy for their extreme values of $\alpha_{EUV}$.  The hardest EUV spectrum is that of
HE~2347-4342 with an $\alpha_{EUV}$ of $+0.56$.  The conspicuously hard UV spectrum, which results in
HE~2347-4342 being quite bright far into the EUV, has made this object a favorite for studying 
\ion{He}{2} \lya\ absorption in the IGM \citep{rkwg+97, ksoz+01}.  Two other
objects have quite soft EUV spectra, the radio-loud QSO MC~1146+111 and the radio-quiet QSO
TON~34.  As can be seen in Figure~\ref{fig:alphalum}, MC~1146+111 is by far the least luminous QSO
for which we measure $\alpha_{EUV}$, so it is perhaps not surprising that it is an outlier, 
particularly given the suggestion of a luminosity dependence for radio-loud objects.
TON~34 is quite extreme, with a measured $\alpha_{EUV}$ of $-5.29$.  The object appears quite normal
in the optical \citep*{ssb88}.  Data from the International Ultraviolet Explorer seem to confirm this
continuum shape down to 2200 \AA, where the HST data end, although it appears that the spectrum may 
flatten at shorter wavelengths \citep*{tbg94}.  We note that the QSO UM~675 has a very similar 
continuum shape to TON~34, although it did not quite meet our stipulated requirements to measure 
$\alpha_{EUV}$.  These objects may be affected by intrinsic reddening by dust in the vicinity of
the QSO.

\begin{figurehere}
\centerline{\psfig{file=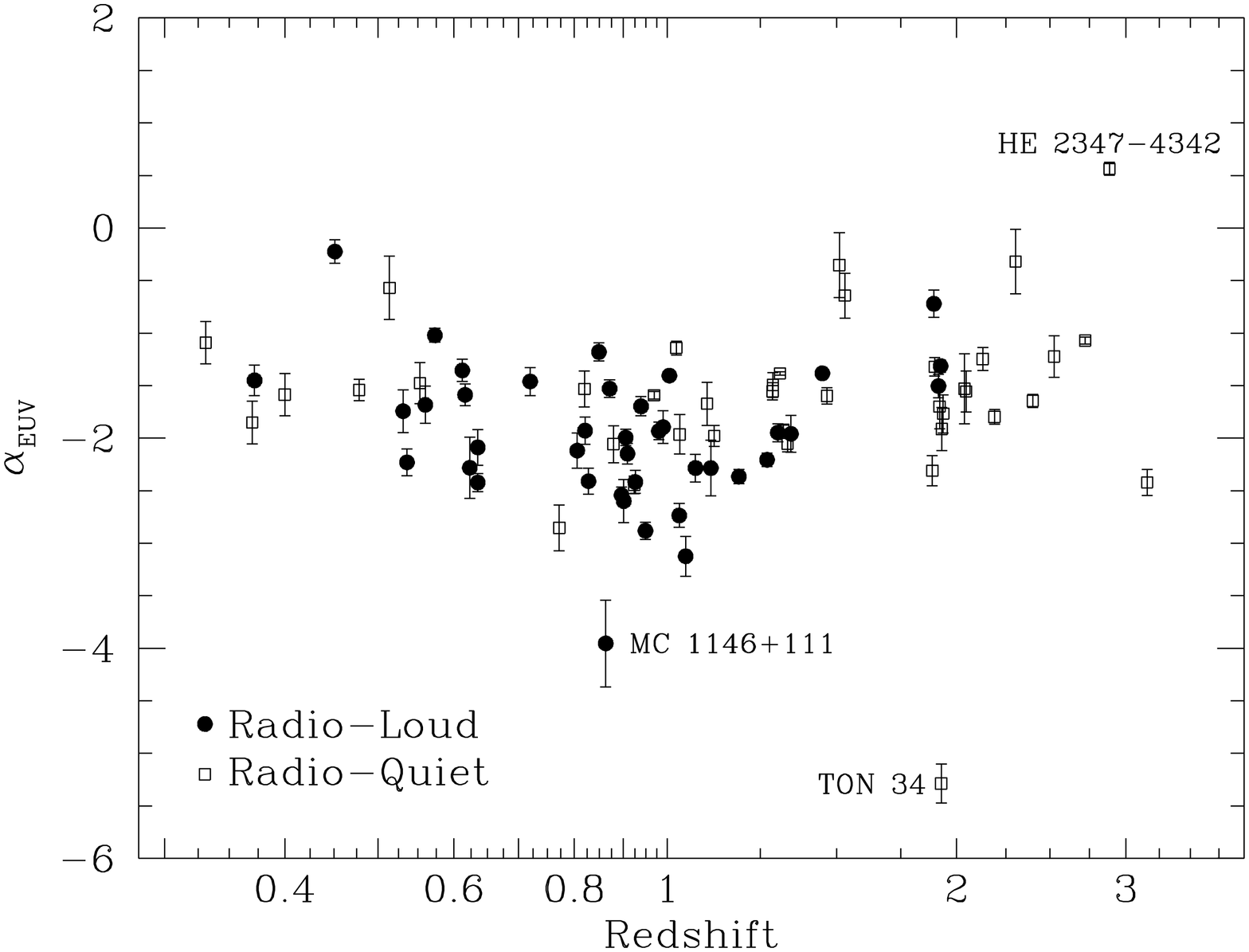,angle=-90,width=9cm}}
\caption{Individual EUV power-law indices for all objects for which it
could be measured, plotted against redshift.\label{fig:alphared}}
\end{figurehere}
\vspace{0.2cm}

\begin{figurehere}
\centerline{\psfig{file=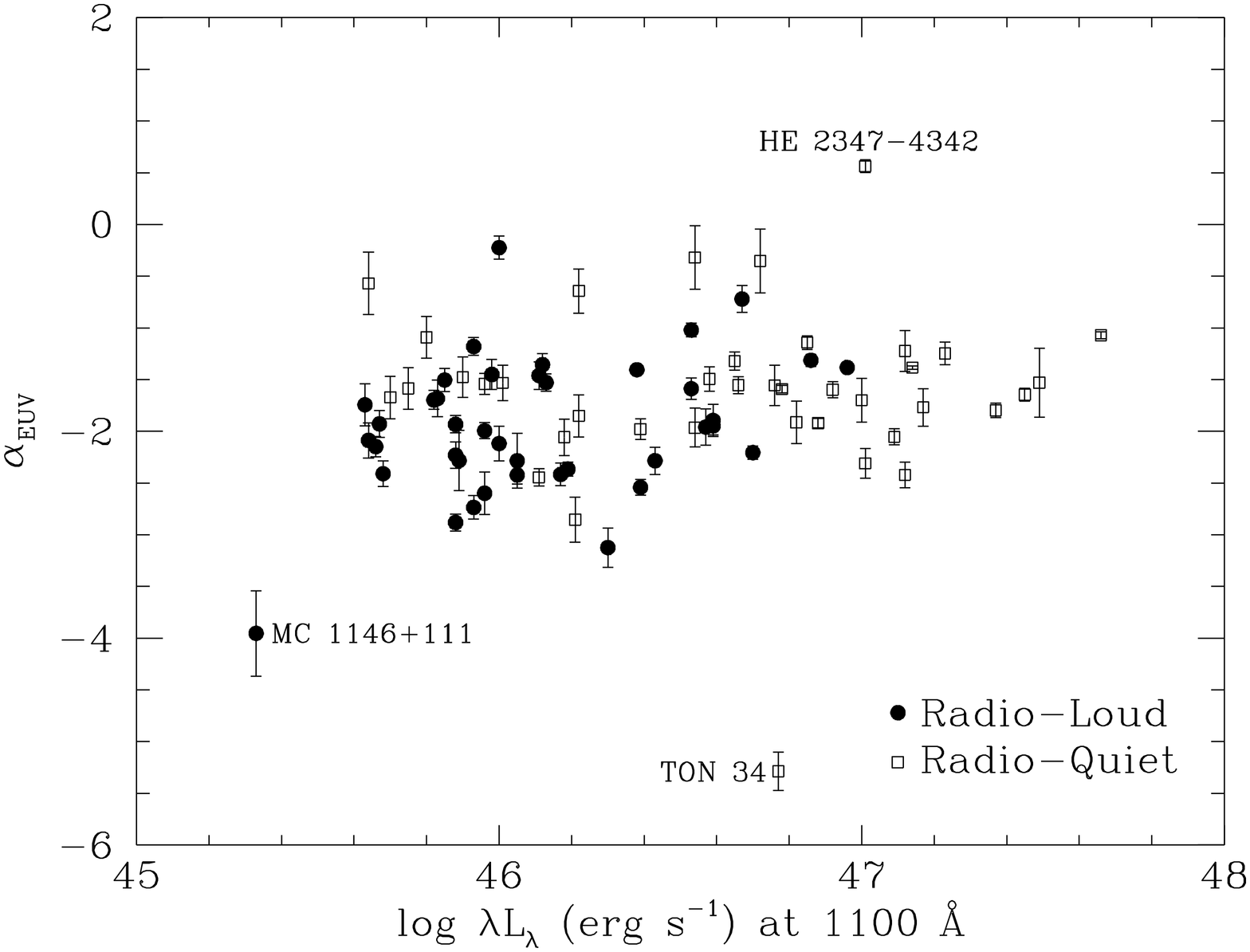,angle=-90,width=9cm}}
\caption{Individual EUV power-law indices for all objects for which it
could be measured, plotted against monochromatic luminosity at 1100 \AA.\label{fig:alphalum}}
\end{figurehere}
\vspace{0.2cm}

For the objects for which we measured $\alpha_{EUV}$, we have also measured
$\alpha_{NUV}$ when possible.  Because very few objects have HST data both below 900 \AA\
and above 2000 \AA, we cannot fit $\alpha_{NUV}$ in exactly the same way as the composite.
To define the fit better, we add additional fit windows:  1315--1325, 1350--1365, 1450--1470,
1760--1840.  We obtain fits of $\alpha_{NUV}$ for 11 radio-quiet and 19 radio-loud objects,
plotted against $\alpha_{EUV}$ in Figure~\ref{fig:euvnuv}.  Clearly, there is no correlation
between $\alpha_{NUV}$ and $\alpha_{EUV}$; the significance of any correlation from the
Spearman rank-order test is 0.055.

In light of a lack of correlation of $\alpha_{NUV}$ with $\alpha_{EUV}$, we return to the
subject of bias in our sample; specifically, whether our sample is a representative one in
the EUV.  The objects for which we measure $\alpha_{EUV}$, by definition, are at redshifts
that place the EUV continuum in the observed-frame UV.  These objects are generally selected
for UV spectroscopy by their brightness in the optical, or the rest-frame NUV region.  Thus,
if there is a bias, one would expect it to be primarily towards harder $\alpha_{NUV}$, as these
objects will be brighter in the observed UV.  However, since Figure~\ref{fig:euvnuv}
suggests no correlation, we conclude that any bias in $\alpha_{NUV}$ should not translate
into a bias in $\alpha_{EUV}$, and thus our sample is likely to be representative of the 
overall population, except at the shortest wavelengths (see \S\ref{sec:comp}).

\begin{figurehere}
\centerline{\psfig{file=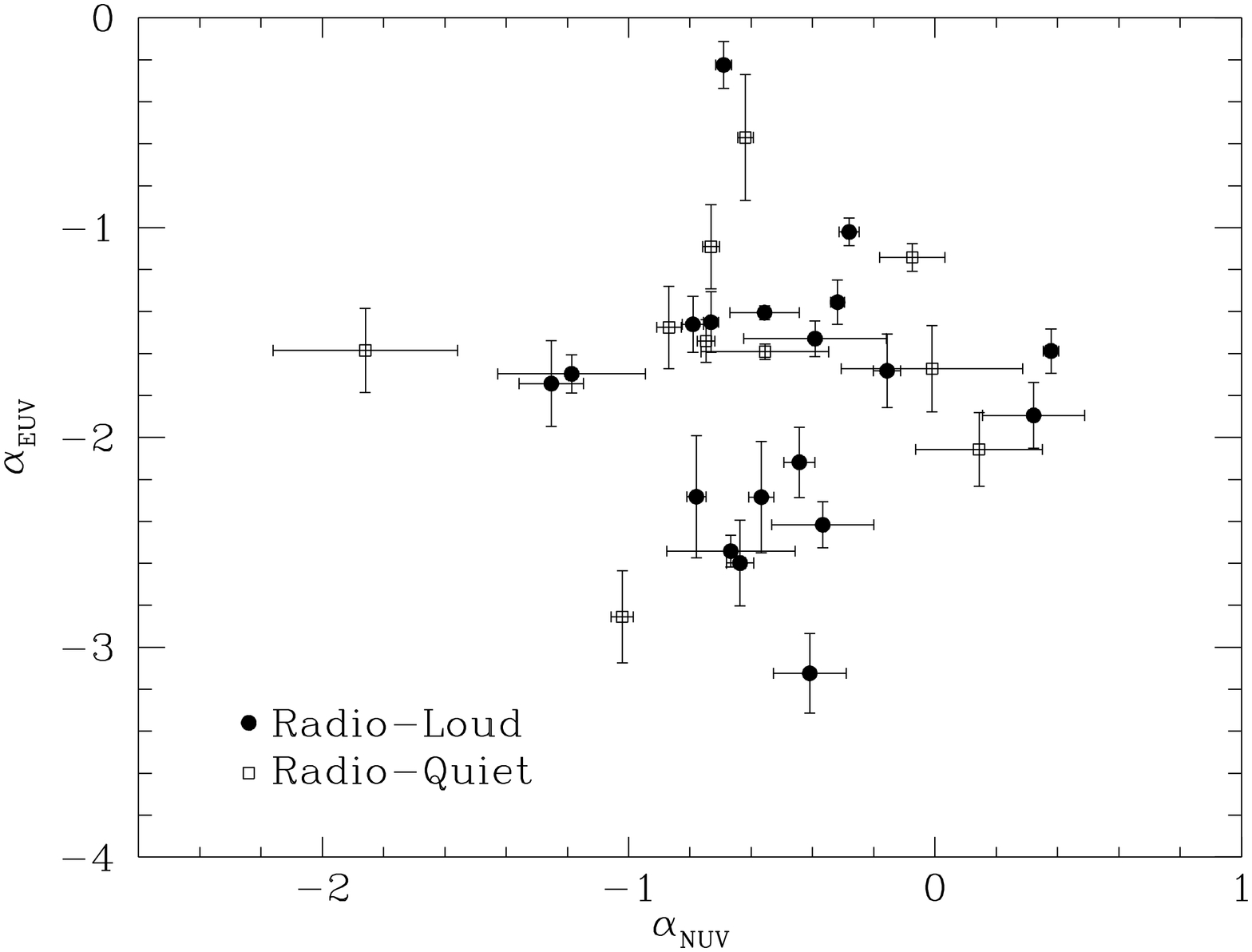,angle=-90,width=9cm}}
\caption{Individual EUV power-law indices vs.\ NUV power-law indices, for 
all objects for which both could be measured.\label{fig:euvnuv}}
\end{figurehere}
\vspace{0.2cm}

\section{DISCUSSION\label{sec:discuss}}
\subsection{Optical Composites\label{sec:optcomp}}
As a basis for comparison of our work to optical composites, we compare our radio-quiet
composite to the SDSS composite \citep{vand+01}.  For our purposes, differences among the 
comprehensive optical composites are small and have been studied by other authors
\citep{btbg+00, vand+01}.  \citet{vand+01} quote a power-law index in their median 
composite between \lya\ and
4000 \AA, equivalent to what we call $\alpha_{NUV}$, of $-0.46$, significantly harder
than our value of $-0.71$.  This can be partially explained by the way in which this
value is measured.  Because we do not have data to such long wavelengths as the SDSS
composite, the longest wavelength windows we use to estimate the continuum are
at 1975--2000 and 2150--2200 \AA, much shorter than the wavelengths spanned by the SDSS composite.
In our composite these appear to be free of emission lines.
However, a longer continuum baseline as is available for the SDSS composite suggests that
even these windows are contaminated by \ion{Fe}{2} emission, which would cause us to infer
a softer continuum.  If we measure the spectral index of the SDSS composite in the same
way as we have done, we get $\alpha_{NUV} = -0.53$, partially explaining the discrepancy.

The remainder of the difference must come from a genuine difference between the spectra.
In Figure~\ref{fig:hstsdss}, we plot our radio-quiet composite along with the SDSS 
composite, with the two normalized to the same flux at 1450 \AA.  The bottom panel shows
the ratio.  Clearly, there is a slight slope to the ratio of the spectra, confirming
that a genuine difference exists.  This difference is likely due to the fact that at
any given wavelength, the mean redshift of objects contributing to our composite is
significantly less than for the SDSS composite.  The difference in $\alpha_{NUV}$ 
can thus be attributed to both a true evolution of QSOs to softer spectra at
lower redshift as well as an observational bias towards detecting QSOs with harder
spectra at high redshift \citep{fran93}.

\begin{figurehere}
\centerline{\psfig{file=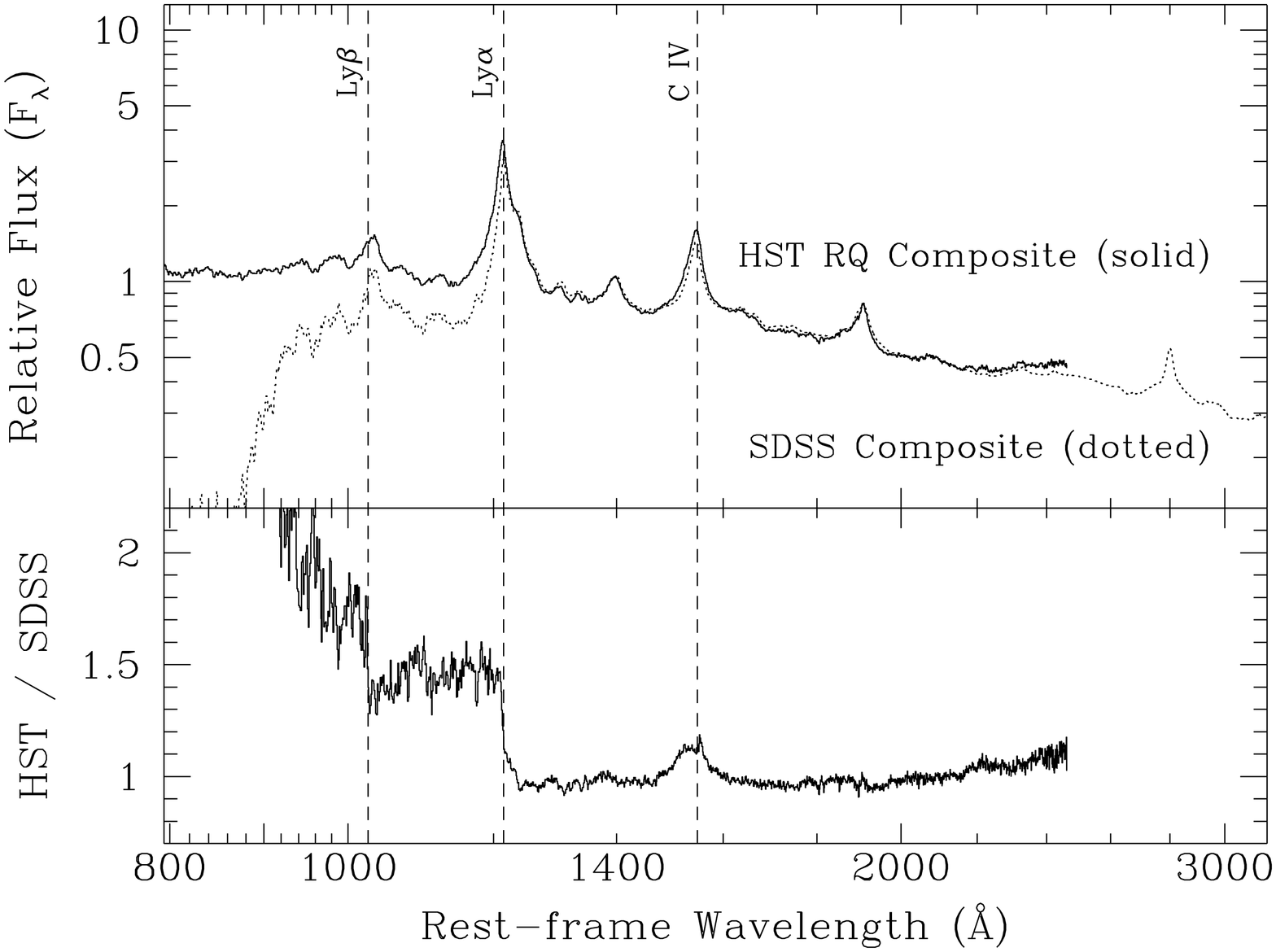,angle=-90,width=9cm}}
\caption{Above, our HST RQ composite compared to the SDSS composite, where the 
two have been normalized to have the same flux at $\sim 1450$ \AA.  Below, the ratio of the
HST composite to the SDSS composite.  The sharp features at \lya\ and \lyb\ are due to 
intervening Lyman-line absorption.  Excess \ion{C}{4} emission is 
evident.\label{fig:hstsdss}}
\end{figurehere}
\vspace{0.2cm}

Only two other major differences are evident in Figure~\ref{fig:hstsdss}.  The first is
the discontinuities at the positions of \lya\ and \lyb\ due to intervening Lyman line opacity.
Our composite does not show these features because we have corrected for line opacity, but 
even if we had not the discontinuities would be much weaker, owing to the much lower
density of intervening absorbers at lower redshift.  The other obvious feature is an excess
of \ion{C}{4} emission in our composite relative to the SDSS.  This is likely due to the known 
tendency of high-ionization emission lines to have lower equivalent widths at higher 
luminosity, the so-called Baldwin effect \citep{bald77}.

\subsection{The EUV / X-Ray Continuum}
For radio-quiet QSOs, our results are in excellent agreement with the existence of a common
power-law continuum describing the entire EUV / soft X-ray spectral energy distribution from 
$\sim 10$~eV to 2~keV.  The ratio of optical to X-ray flux is conventionally represented by the 
effective power-law index $\alpha_{ox}$, defined as $\alpha_{ox} = (l_o - l_x) / 2.605$, where 
$l_o$ and $l_x$ are the common logarithms of the monochromatic fluxes in units of \nuflux\ at 
2500 \AA\ and 2 keV, respectively.  Assuming the continuum shape of our composite to convert from
$\lambda L_{\lambda}$ at 1100 \AA\ to $l_o$, the QSOs in our sample fall in the range 
$30 \lesssim l_o \lesssim 32.5$.  Typical mean values of $\alpha_{ox}$ from samples of QSOs at these
luminosities tend to fall in the range 1.5--1.6 
(\citealt{anma87}; \citealt*{awm95}; \citealt{gsah+95}; \citealt{ybsv98}).
Assuming the NUV spectral shape of our composite and extrapolating the composite EUV continuum 
with $\alpha_{EUV}$ of $-1.57 \pm 0.17$ to 2 keV, we would 
expect an $\alpha_{ox}$ of $1.50 \pm 0.15$, consistent with the measurements of $\alpha_{ox}$.
Although we find no correlation of $\alpha_{EUV}$ with luminosity for 
radio-quiet objects (\S\ref{sec:indiv}), our limits on the slope are consistent with the 
slight dependence of $\alpha_{ox}$ on luminosity ($d\alpha_{ox} / dl \approx 0.1$) found in
X-ray studies of optically-selected AGN \citep{krca85, awm95}.  Our finding of no evolution with
redshift is also consistent with these studies.
In addition, our EUV continuum shape matches well with the typical soft X-ray continuum slope 
of low-redshift
QSOs.  \citet{lfew+97} find a mean $\alpha_x$ of $1.72 \pm 0.09$ for a sample of 19 $z < 0.4$ 
radio-quiet
QSOs, while \citet{ybsv98} find a mean $\alpha_x$ of $1.58 \pm 0.05$ for a much larger sample of
146 $z < 0.5$ radio-quiet QSOs.  Thus both the typical continuum shape and luminosity of 
radio-quiet QSOs in the soft X-ray region are in good agreement with an extrapolation of our
results to higher energies.

The situation is more complicated for radio-loud QSOs.  While an extrapolation of our EUV 
continuum with $\alpha_{EUV} = -1.96\pm 0.12$ yields $\alpha_{ox} = 1.84 \pm 0.10$, 
data show that radio-loud QSOs are much more luminous in the soft X-ray band than this extrapolation.
\citet*{bys97} find a mean value of $\alpha_{ox}$ of 1.24 for flat spectrum sources and
1.33 for steep spectrum sources.  They also find typical power-law indices in the soft X-ray
$\sim -1.0$, although this is found to depend on the properties of the radio emission.
Thus radio-loud QSOs are both much too luminous and too flat in the soft X-ray to be explained as
an extrapolation of the EUV emission, and there must be some significant change in spectral shape
somewhere between $\sim 50$ eV and a few tenths of a keV, corresponding to the limits of the 
studied portions of the rest-frame EUV and soft X-ray bands.

The apparent existence of a common EUV / X-ray continuum in radio-quiet QSOs 
results in a return of the problem for which the EUV bump of \citet{mafe87} was hypothesized;
specifically, the observed strength of \ion{He}{2} \lam 1640 emission.  Assuming 
$\alpha_{EUV} = -2$, \citet*{kfb97} pointed out that it is difficult to generate the observed amount
of \ion{He}{2} \lam 1640 emission with such a soft ionizing spectrum.  In terms of an ensemble
property of radio-quiet QSOs, our currently favored value of $\alpha_{EUV} = -1.57$
lessens this problem somewhat by providing nearly twice as many photons at 4 Ryd.  Scaling the
calculations of \citet{kfb97}, assuming as they did a covering fraction of 10\%, we 
would expect a \ion{He}{2} \lam 1640 EW of $\sim$1--2 \AA, where this range takes into account some
uncertainty in the modeling and our error on $\alpha_{EUV}$.  This estimate in fact agrees well with our
measured value of 1.3 \AA.  However, in our decomposition of the \ion{C}{4} blend we have made
a conservative measurement of the \ion{He}{2} \lam 1640 emission, fitting only the well-defined 
core of the line above the broad $\sim 1600$ \AA\ feature, and our value is low in comparison to 
typical values measured for individual objects using more detailed methods, even 
those obtained from many of the same spectra that comprise our sample \citep{lbjs+95, msdc+96}.
If any significant amount of flux from the $\sim 1600$ \AA\ feature 
belongs to a broader component of \ion{He}{2} \lam 1640, we may be underestimating the flux
by up to a factor of several, which cannot be easily explained with our proposed ionizing continuum.

\subsection{Ionization of the Intergalactic Medium}
The observation of strong absorption by \ion{He}{2} \lya\ \lam 304 \AA\ in high-redshift QSO
spectra (\citealt{jbdg+94}; \citealt*{dkz96}; \citealt{rkwg+97}) have shown that the bulk of the IGM 
is in a highly-ionized state.  This is generally believed to be due to photoionization by integrated
EUV radiation from QSOs and/or hot, massive stars.
Because radio-quiet QSOs are much more common than their radio-loud counterparts 
\citep[e.g.,][]{srwe80,cops81}, the radio-quiet QSO spectral energy distribution is generally
taken as representative of the QSO population, so we use our radio-quiet value of 
$\alpha_{EUV} = -1.57\pm 0.17$ to compare to the literature.  Many models of
the ionization of the IGM (\citealt{hama96}; \citealt*{fgs98}, \citealt*{mhr99}) 
assume the dominant
source of the metagalactic ionizing is QSOs, although the latter two articles discuss the
contribution of stars \citep[see also][]{mash96, hmkh01}.  These models assume that the QSO
ionizing radiation is emitted as a pure power law but is subsequently reprocessed
through absorption and reemission by the intergalactic gas.

\citet{ksoz+01} have recently completed a detailed analysis of a
Far Ultraviolet Spectroscopic Explorer (FUSE) spectrum of HE~2347-4342 that provides data on
the \ion{He}{2} \lya\ forest for $2.3 < z < 2.7$.  They attribute the \ion{He}{2} absorption 
entirely to discrete absorbers
and find a large range in the ratio of \ion{He}{2} to \ion{H}{1} column densities ($\eta$) of these 
absorbers, from $\eta = 1$ to $\eta > 1000$, with a mean value of $\langle \eta \rangle = 78 \pm 7$.
Comparing to the models of \citet{fgs98}, $\alpha_{EUV} = 1.57 \pm 0.17$ would lead to values
of $\eta$ in the range 25--60.  This is consistent with the lower values in the observed range
of $\eta$, but it cannot explain the highest values observed.  However, as is evident in 
Figure~\ref{fig:alphahist}, individual radio-quiet QSOs show a wide spread in $\alpha_{EUV}$
with an RMS deviation of 0.86.  Deviations of this magnitude from the mean would correspond to
a range of $\eta$ from $\eta < 10$ to $\eta > 100$ according to the \citet{fgs98} models.
Although there is considerable uncertainty in the modeling due to the many input parameters
involved, and the uncertainty on our measurement of $\alpha_{EUV}$ is substantial, it is
possible that the typical QSO spectrum is somewhat too hard to explain the highest values of 
$\eta$ ($> 1000$) observed in the IGM, and an additional source of hydrogen ionizing photons, such
as starburst galaxies, may be necessary to soften the metagalactic ionizing radiation.

\section{SUMMARY\label{sec:summary}}

We have studied the UV spectral properties of QSOs using 332 HST spectra of 184 $z > 0.33$ QSOs.
This sample is nearly a factor of two larger than that of \zh, with the largest improvement
coming in the 600--800 \AA\ range, where our composite is composed of up to five times as many
spectra as that of \zh.  We confirm the result of \zh\ that the UV spectral continuum can be 
approximated by a broken power law with the break in the vicinity of \lya.  However, we find
a somewhat harder continuum than \zh.  For the overall composite, we find a continuum power-law
index between 500 \AA\ and the break of $\alpha_{EUV} = -1.76 \pm 0.12$.  For the subset of 
exclusively radio-loud objects, we find  $\alpha_{EUV} = -1.96 \pm 0.12$; for the radio-quiet
subset, $\alpha_{EUV} = -1.57 \pm 0.17$.  Based on the bootstrap technique of resampling
the spectra to generate the likely distributions of the measured values of $\alpha_{EUV}$, 
we find that the radio-quiet QSOs have harder EUV spectra than radio-loud QSOs with 
98.6\% confidence.

To verify that the composite is representative of the sample spectra, we measure $\alpha_{EUV}$
for as many individual objects as possible in the same manner as for the composite.  We find
median values of $\alpha_{EUV}$ of $-1.59$ for the radio-quiet QSOs and $-1.95$ for the radio-loud
QSOs, in excellent agreement with the composites.  We apply a two-population Kolmogorov-Smirnov 
test to the two distributions of $\alpha_{EUV}$, with the result that the radio-loud and radio-quiet
values were not drawn from the same parent distribution with 98.6\% confidence.
Also, we use these data on individual objects to search for dependencies of $\alpha_{EUV}$ on
luminosity or redshift.  We find no evidence for evolution of $\alpha_{EUV}$ with redshift for
either radio-loud or radio-quiet QSOs.  We also find no evidence for a correlation between
 $\alpha_{EUV}$
and luminosity for radio-quiet QSOs, though the limits are consistent with the slight increase of 
$\alpha_{ox}$ with increasing luminosity found in some X-ray studies.  The radio-loud QSOs 
indicate a possible correlation with luminosity, 
with the trend being towards harder spectra with increasing luminosity.  A Spearman rank-order
correlation test indicates that the correlation is real with 90.9\% confidence.

An extrapolation of our radio-quiet composite to higher energies is in good agreement with X-ray
data, indicating that it is plausible to represent the entire typical ionizing continuum from 
$\sim 10$ eV to $\sim 2$ keV by a single power law.  However, radio-loud QSOs are in general both
flatter and more luminous in soft X-ray regime than an extrapolation of our composite, suggesting
that there should be a change in spectral shape between the EUV and soft X-ray regions.  The fact
that radio-quiet QSOs may be represented by a single power law with spectral index 
$\alpha_{EUV} = -1.57 \pm 0.17$ is roughly in agreement with models of the IGM
being photoionized by integrated QSO radiation as filtered by intergalactic material,
given the uncertainty of our result and those of the models.  However, as recent results 
show that at least some regions of the IGM are photoionized by even softer spectra, 
it is possible that an additional source of soft ionizing radiation, such as hot, massive stars, 
may be necessary to fully explain the ionization state of the IGM.

\acknowledgements
This research has been supported by grant
AR-07977.01-96A from the Space Telescope Science Institute, which is operated 
by the Association of Universities of Research in Astronomy, Inc.,
under NASA contract NAS5-26555.

\end{document}